\documentclass[11pt]{article}
\usepackage{verbatim,amsmath,amssymb}
\usepackage{epsfig,float,color}
\usepackage{epstopdf}
\usepackage{geometry}
\usepackage{setspace}
\usepackage[numbers]{natbib}
\usepackage{hyperref}
\usepackage[utf8]{inputenc}
\usepackage{graphicx}
\usepackage{flushend}
\usepackage[font={footnotesize,sf,bf}]{caption}
\usepackage[font={footnotesize,sf,bf}]{subcaption}
\usepackage{footnote}
\usepackage{multicol}
\usepackage{multirow}
\usepackage{framed}
\usepackage{bm}
\usepackage{soul}
\definecolor{darkblue}{rgb}{0,0,1}
\hypersetup{pdftex=true, colorlinks=true, breaklinks=true, linkcolor=darkblue, menucolor=darkblue, pagecolor=darkblue, citecolor=darkblue, urlcolor=darkblue}

\begin{document}

\begin{center}
	\Large{\bf{Computational synthesis of large deformation compliant mechanisms undergoing self and mutual
			contact}}\\
	
\end{center}

\begin{center}
	\large{Prabhat Kumar\,$^{a,}\,^{b,}$\footnote{corresponding author, email: p.kumar-3@tudelft.nl}, Anupam Saxena\,$^{a},$ and Roger A. Sauer$\,^{c}\!\!$} \\
	\vspace{4mm}
	
	\small{\textit{$^a$Mechanical Engineering, Indian Institute of Technology Kanpur, Kanpur, 208016, India}}
	
	\small{\textit{$^b$Department of Precision and Microsystems Engineering, Faculty of
			3mE, Delft University of Technology, Mekelweg 2, 2628 CD, Delft, The
			Netherlands}}
	
	\small{\textit{$^c$Aachen Institute for Advanced Study in Computational Engineering Science (AICES), RWTH Aachen University, Templergraben 55, Aachen 52056, Germany}}
	
	\vspace{4mm}
	
	Published\footnote{This pdf is the personal version of an article whose final publication is available at \href{http://mechanicaldesign.asmedigitalcollection.asme.org/article.aspx?articleid=2696698}{http://asmedigitalcollection.asme.org}} 
	in \textit{Journal of Mechanical Design}, 
	\href{http://mechanicaldesign.asmedigitalcollection.asme.org/article.aspx?articleid=2696698}{DOI: 10.1115/1.4041054} \\
	Submitted on 28.~May 2018, Revised on 29. July 2018, Accepted on 3.~August 2018
	
\end{center}

\vspace{3mm}
\rule{\linewidth}{.15mm}
{\bf Abstract}
	Topologies of large deformation Contact-aided Compliant Mechanisms (CCMs), with self and mutual contact, exemplified via path generation applications, are designed using the continuum synthesis approach. Design domains  are parameterized using honeycomb tessellation. Assignment of material to each cell, and generation of rigid contact surfaces, are accomplished via suitably sizing and positioning negative circular masks. To facilitate contact analysis, boundary smoothing is implemented. Mean value coordinates are employed to compute shape functions, as many regular hexagonal cells get degenerated into irregular, concave polygons as a consequence of
	boundary smoothing. Both, geometric and material nonlinearities are considered in the finite element analysis. The augmented Lagrange multiplier method in association with an active set strategy is employed to incorporate both self and mutual contact. CCMs are evolved using the stochastic hill climber search. Synthesized contact-aided compliant continua trace paths with single and importantly, multiple kinks and experience multiple contact interactions pertaining to both self and mutual contact modes. 
	\\
	

\vspace{-4mm}
\rule{\linewidth}{.15mm}


\section{INTRODUCTION}\label{I}
Contact-aided Compliant Mechanisms (CCMs) transfer energy, force, and motion in a desired manner via large deformation of their flexible members that experience mutual and/or self-contact. Self-contact occurs when a body comes into contact with itself, while in mutual contact, the body interacts with neighboring objects (Fig. \ref{fig:fig_1}). Compliant mechanisms, in general, yield smooth output responses if  the material model is continuous and/or buckling in their members is not permitted \cite{mankame2004topology,kumar2016synthesis}. In contrast, contact constraints alter deformation characteristics of compliant mechanisms instantly thereby helping achieve nondifferentiability in their output responses. However, such constraints introduce strong boundary nonlinearities when contact pairs are not known a priori \cite{wriggers2006computational}. These nonlinearities become even more pronounced when boundaries evolve, e.g, in topology optimization. In addition, large deformation in flexible branches of a CCM requires consideration for geometrical and/or material nonlinearities within the synthesis approach. Several other challenges, e.g., contact-pair detection and dynamic mesh handling (temporary removal of non-existing cells for contact analysis) need to be addressed when synthesizing CCMs especially via topology optimization. 

Numerous approaches exist  to synthesize optimal topologies of compliant mechanisms for different applications \cite{howell2001compliant}. These approaches extremize the objectives stemming from a combination of flexibility property (e.g., output displacements) and strength/stiffness measure (e.g., stress constraints/strain energy) of the mechanisms. Ananthasuresh et al. \cite{ananthasuresh1994methodical} employed the homogenization approach with an optimality criterion to minimize the linearly weighted objective involving  output displacement and strain-energy. Nishiwaki et al. \cite{nishiwaki1998topology} and Frecker et al. \cite{frecker1997topological} maximized the ratio of flexibility and stiffness measures. Saxena and Ananthasuresh \cite{saxena2000optimal}  generalized the multi-criteria objective. Sigmund \cite{sigmund1997design} optimized an objective based on mechanical advantage. Saxena and Ananthasuresh \cite{saxena2001topology} and Pedersen et al. \cite{pedersen2001topology} synthesized path generating fully compliant mechanisms  considering nonlinearity in geometry alone. In \cite{saxena2001topology}, line tessellation was used while in \cite{pedersen2001topology}, rectangular mesh was employed. Synthesis approaches in \cite{pedersen2001topology,saxena2001topology,saxena2005synthesis} involved  least-square error objectives. Swan and Rahmatalla \cite{swan2004design} proposed a control based approach to obtain a compliant mechanism which could also trace paths close to the specified path. Ullah and Kota \cite{ullah1997optimal} used Fourier Shape Descriptors (FSDs) \cite{zahn1972fourier}, as the least square objective introduces an unnecessary timing constraint making it difficult to search for an optimal solution. Rai et al. \cite{rai2007synthesis,rai2010unified} employed such an objective to synthesize path generating fully and partially compliant mechanisms with curved beam and rigid truss elements. Saxena \cite{saxena2008material,saxena2011adaptive} presented the Material Masks Overlay Strategy (MMOS) which uses hexagonal cells to discretize the design space, and negative circular masks to decipher material states of each cell. Gradient based optimization was used in \cite{saxena2011topology}. Saxena and Sauer \cite{saxena2013combined} combined zero and first order search approaches to synthesize such mechanisms considering geometrical and material nonlinearities. 

CCMs were introduced by Mankame and Anathasuresh \cite{mankame2002contact}. They synthesized CCMs via topology optimization using frame elements with intermittent rigid contact surfaces \cite{mankame2004topology} and later extended their work to synthesize non-smooth path generating CCMs by considering large deformations, and using an objective based on FSDs \cite{mankame2007synthesis}. Therein, contact locations were prespecified. Reddy et al. \cite{reddy2012systematic} used curved beam elements to parametrize the design space and found contact locations systematically. Tummala et al. \cite{tummala2014design} designed a compliant spine, a CCM joint that is flexible in one direction while rigid in the other, using the multi-objective formulation.  Kumar et al. \cite{kumar2016synthesis} synthesized $C^0$ path generating CCMs with continuum discretization using hexagonal cells where external, rigid contact surfaces were generated automatically. In \cite{kumar2017implementation}, they synthesized such mechanisms with only self contact. Various CCM designs and applications are presented in \cite{cannon2005compliant,moon2007bio,aguirre2008fabrication,mehta2009stress,saxena2013contact,calogero2016dynamic}. As aforementioned, few works exist that address topology design of contact-aided compliant mechanisms. Of those, ones that employ triangular, quadrilateral or hexagonal cell parameterization are rare. To our knowledge, design of CCMs that observe both \textit{self} and \textit{mutual} contact modes at multiple sites has not been addressed yet using continuum topology optimization. 

The notion of contact MMOS is extended herein, for synthesis of Contact Aided Compliant Mechanisms the constituents of which can undergo a number of both, \textit{self} and \textit{mutual} contact modes. CCMs are synthesized with only mutual contact in \cite{kumar2016synthesis}, and only self contact in \cite{kumar2017implementation}. By combining both interaction modes in this paper, as opposed to restricting their nature as in \cite{kumar2016synthesis,kumar2017implementation}, potency of the design space is showcased via CCM examples which can witness multiple contact interactions, both in number and mode, and exhibit comparatively more intricate deformation characteristics, e.g., tracing of a desired output path with multiple kinks.

In particular, the present approach is novel in the following aspects compared to our previous efforts in \cite{kumar2016synthesis,kumar2017implementation}:
\begin{itemize}
	\item Continuum synthesis via contact MMOS of CCMs which experience both, \textit{self} and \textit{mutual} contact at multiple sites, all determined systematically by the proposed algorithm, during their deformation. This was proposed as future work in \cite{kumar2016synthesis}, section 8. In particular,  a CCM that traces a desired, challenging $Z$-path with two non-differentiable sites is synthesized.
	\item Exploring the boundary smoothing scheme further (section \ref{BSMVCF}) and noting that higher number of boundary smoothing steps may lead to \textquoteleft element flipping\textquoteright.\, To assuage notches at the continuum boundaries to facilitate contact analysis, few boundary smoothing steps are adequate.
	\item Formulation of an active set strategy (Table \ref{T1}) to examine active or inactiveness of both, self and/or mutual contact boundary constraints within the Newton-Raphson solution iterations.
	\item 	Presentation of many examples with a variety of desired output paths wherein multiple number and/or modes of contacts are witnessed, as intended. In examples in \cite{kumar2016synthesis}, only a single mutual contact mode is observed. 
	\item Appraising the synthesis approach by varying mesh sizes (section \ref{coarse_fine_mesh}), number of integration points (section \ref{CMPCSTs}), estimating computational costs (section \ref{CMPCSTs}), and demonstrating existence of multiple CCM solutions  for identical design specifications.
\end{itemize}
The remainder of the paper is organized as follows: Section \ref{TOCSG} describes topology optimization with hexagonal cells and negative circular masks, the latter also having the capability to generate contact surfaces. Boundary smoothing (Section \ref{BSMVCF}) is used to subdue jumps of the surface normals. The finite element formulation with contact is briefed, and the active set strategy with self contact search is discussed in Section \ref{CMFWWC}. Friction and adhesion are not considered, though friction \cite{sauer2015unbiased} can be  incorporated in the formulation. In Section \ref{PF}, FSDs objective and hill climber search are presented. Synthesized CCMs are presented in Section \ref{SE}. Lastly, the synthesis approach and examples are discussed and conclusions are drawn in Sections \ref{con} and \ref{closure}, respectively.
\section{TOPOLOGY OPTIMIZATION AND CONTACT SURFACE GENERATION} \label{TOCSG}
\begin{figure}
	\centering
	\includegraphics[scale =0.75]{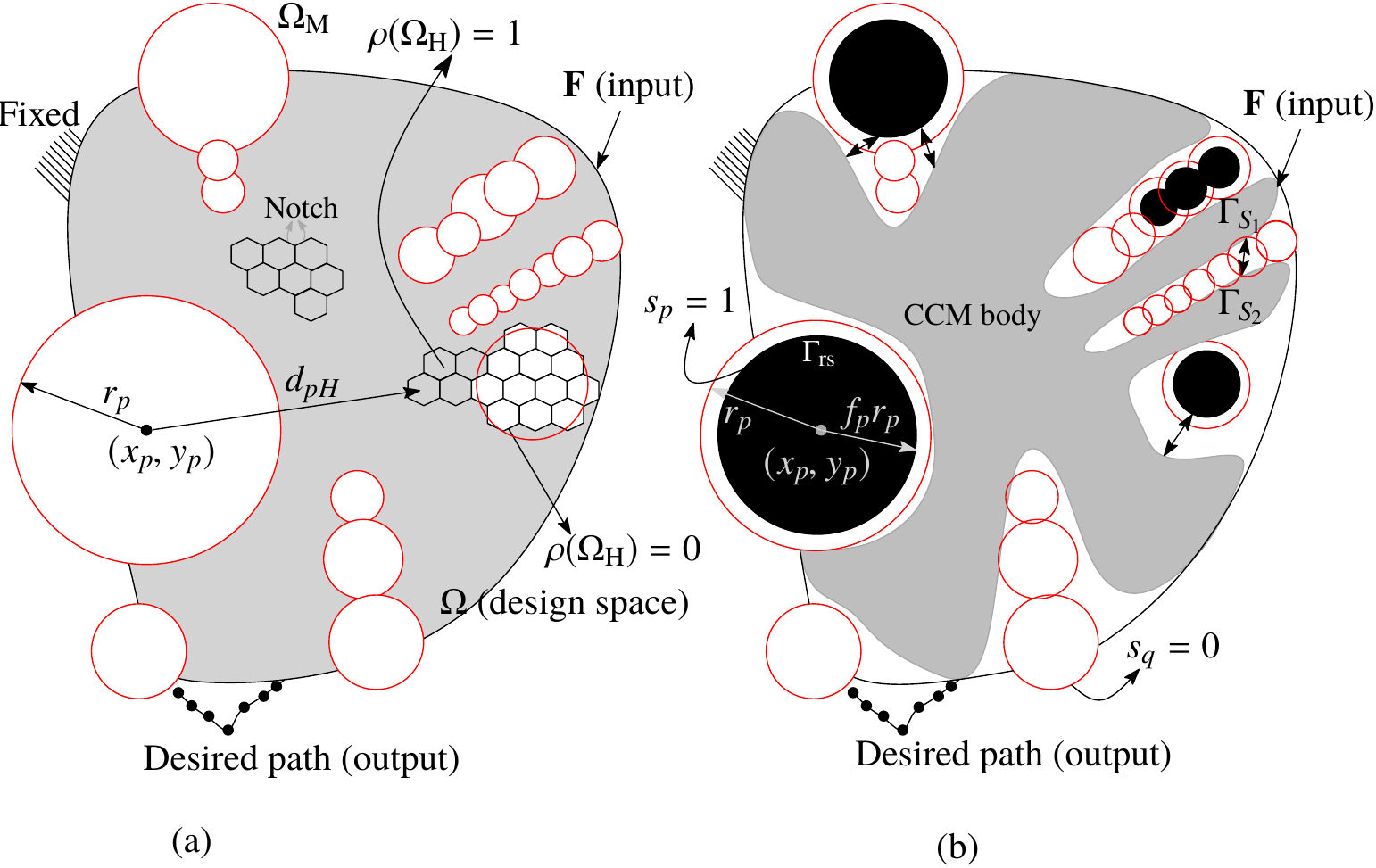}
	\caption{A set of hexagonal cells $\Omega_\mathrm{H}$ discretize the design space $\Omega$. Negative circular masks $\Omega_\mathrm{M}$ (circles) superposed on $\Omega$ help determine the material state of $\Omega_\mathrm{H}$. Five variables ($x_p,\,y_p,\,r_p,\,s_p,\,f_p $) define each mask. $(x_{p}, y_{p})$ and $r_{p}$ are center coordinates and the radius of the $p^\mathrm{th}$ mask. $s_{p}=1$ (figure b) implies a rigid contact surface (dark, filled circular regions) of radius $f_{p}r_{p}$ is generated within the $p^\mathrm{th}$ mask while with $s_{p}=0$ (figure a), no contact surface is generated. $\rho(\Omega_\mathrm{H})=0$ ($\Omega_H \subset$ any $\Omega_M$) implies a void material state while $\rho(\Omega_\mathrm{H})=1$ indicates full material state. Rigid contact surfaces can interact with the continuum. In addition, surfaces (e.g., $\Gamma_{s_{1}}$ and $\Gamma_{s_{2}}$) of the continuum may interact in self contact mode.}
	\label{fig:fig_1}
\end{figure}
Various features of topology optimization with hexagonal cells ($\Omega_\mathrm{H}$, \cite{saxena2003honeycomb, langelaar2007use, saxena2007honeycomb, talischi2009honeycomb, talischi2012polytop}) and negative circular masks ($\Omega_\mathrm{M}$) are described in \cite{saxena2008material, saxena2010adaptive, saxena2011topology, saxena2013combined}. Hexagonal cells provide \textit{edge-connectivity} between any two contiguous cells within the parameterized design. Consequently, point-connections and checkerboards patterns are automatically alleviated \cite{saxena2008material,langelaar2007use,talischi2009honeycomb,talischi2012polymesher,talischi2010polygonal}. One can still observe layering/islands and V-notches on the boundary in the final continua \cite{saxena2011topology}. Blurred boundary also persists with gradient search \cite{saxena2011topology,kumar2015topology}. V-notches (Fig. \ref{fig:fig2a}) render jumps in the boundary normals, which is not desirable in contact analysis \cite{corbett2014nurbs}. To subdue these, boundary smoothing \cite{kumarembedded,kumar2015topology} is employed (Fig. \ref{fig:fig2c}, Section \ref{BSMVCF}). 

Material state $\rho(\Omega_\mathrm{H})$ of each cell is determined via negative circular masks ($\Omega_\mathrm{M}$) which act as material sink \cite{saxena2011topology,saxena2013combined}. Ideally, $\rho(\Omega_\mathrm{H})$ should either be 0 or 1. In the Material Mask Overlay Strategy, $\rho(\Omega_\mathrm{H}) =0$ is set when the centroid $\Omega_\mathrm{H}^\mathrm{c}$ of a cell $\Omega_\mathrm{H}$ is within an overlaying mask and $\rho(\Omega_\mathrm{H}) =1$ is set when $\Omega_\mathrm{H}^\mathrm{c}$ is not enclosed within any mask (Fig. \ref{fig:fig_1}a). Thus, one defines $\rho(\Omega_\mathrm{H})$ as
\begin{equation}
\rho(\Omega_\mathrm{H})=
\begin{cases}
0,&\,\text {if}\,\,\Omega_\mathrm{H}^\mathrm{c}  \subset\,\text{any}\,\,\Omega_\mathrm{M} \\
1,&\,\,\, \text{otherwise}
\end{cases}
\end{equation}
\noindent All cells with $\rho(\Omega_\mathrm{H}) =1$ constitute a potential candidate continuum ($ \mathrm{ T}^0_\mathrm{H}$) for the CCM (Fig. \ref{fig:fig_1}b), i.e., 
\begin{equation}
\mathrm{ T}^0_\mathrm{H}=\{\Omega_\mathrm{H}|\Omega_\mathrm{H} \in \Omega;\,\,\,\Omega_\mathrm{H}^\mathrm{c} \notin\,\text{any}\hspace{1mm} \Omega_\mathrm{M} \,\Rightarrow \,\rho(\Omega_\mathrm{H})=1\}.
\end{equation}

\subsection{Negative Circular Masks and Mutual Contact Surfaces}
\label{NCMnew}
The $p^\mathrm{th}$ mask is defined via its center coordinates ($x_p,\,y_p$) and radius $r_p$. As in \cite{kumar2016synthesis}, masks are also used to generate rigid contact surfaces $\Gamma_\mathrm{rs}$ (Fig. \ref{fig:fig_1}b), which necessitates the use of two additional variables, i.e, $s_p\,\text{and}\,f_p$. $s_p$ is strictly 0 or 1, and $0\le f_p\le 1$. $s_p = 1$ implies that a rigid contact surface $\Gamma_\mathrm{rs}$ of radius $f_pr_p$ is generated within the mask (Fig. \ref{fig:fig_1}b). Mask variables are evolved stochastically via the hill climber search. With $M$ masks, the design vector $\mathbf{v}$ contains $5M$ variables ($\mathbf{v} = \{x_p, y_p, r_p, s_p, f_p, ... \}, p = 1, ..., M$) when both \textit{self} and \textit{mutual} contact modes are permitted. The number of design variables relate to the number of masks used and is independent of the	number of hexagonal cells in the tessellation. While boundary smoothing is not a necessity, it is still desirable to achieve better convergence in contact analysis.

\subsection{Boundary Smoothing} \label{BSMVCF}
\begin{figure}[h!]
	\centering
	\begin{subfigure}{0.30\columnwidth}
		\centering
		\includegraphics[scale=0.6]{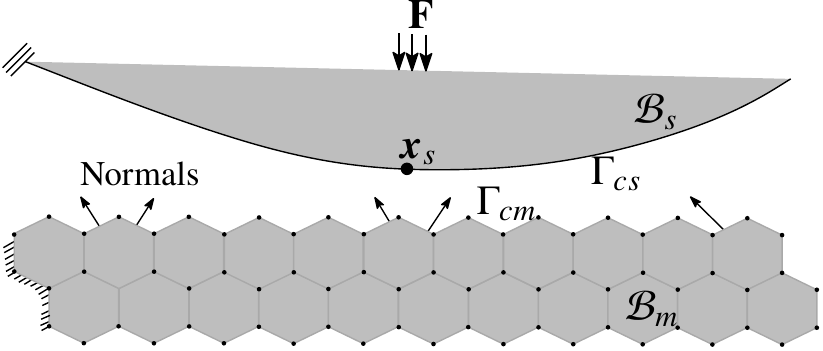}
		\caption{Body $\mathcal{B}_m$ without boundary smoothing}
		\label{fig:fig2a}
	\end{subfigure}
	\qquad
	\begin{subfigure}{0.30\columnwidth}
		\centering
		\includegraphics[scale=0.6]{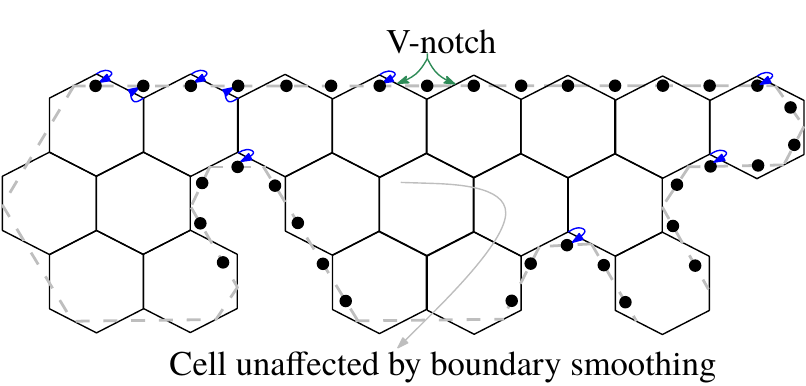}
		\caption{Boundary smoothing scheme}
		\label{fig:fig2b}
	\end{subfigure}
	\begin{subfigure}{0.30\columnwidth}
		\centering
		\includegraphics[scale=0.6]{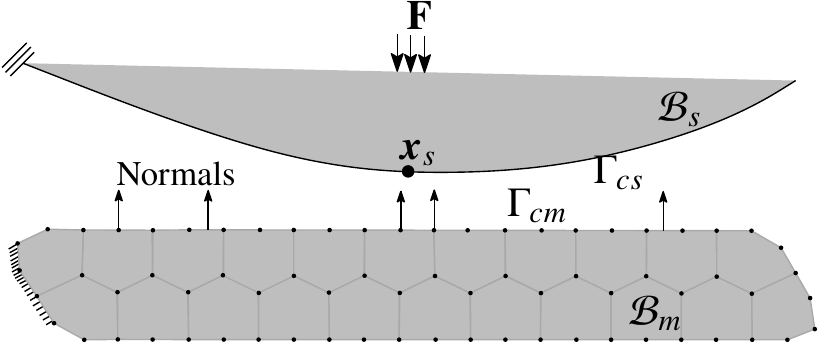}
		\caption{Body $\mathcal{B}_s$ with boundary smoothing; considering $\beta=1$}
		\label{fig:fig2c}
	\end{subfigure}
	\begin{subfigure}{0.90\columnwidth}
		\centering
		\includegraphics[scale=0.70]{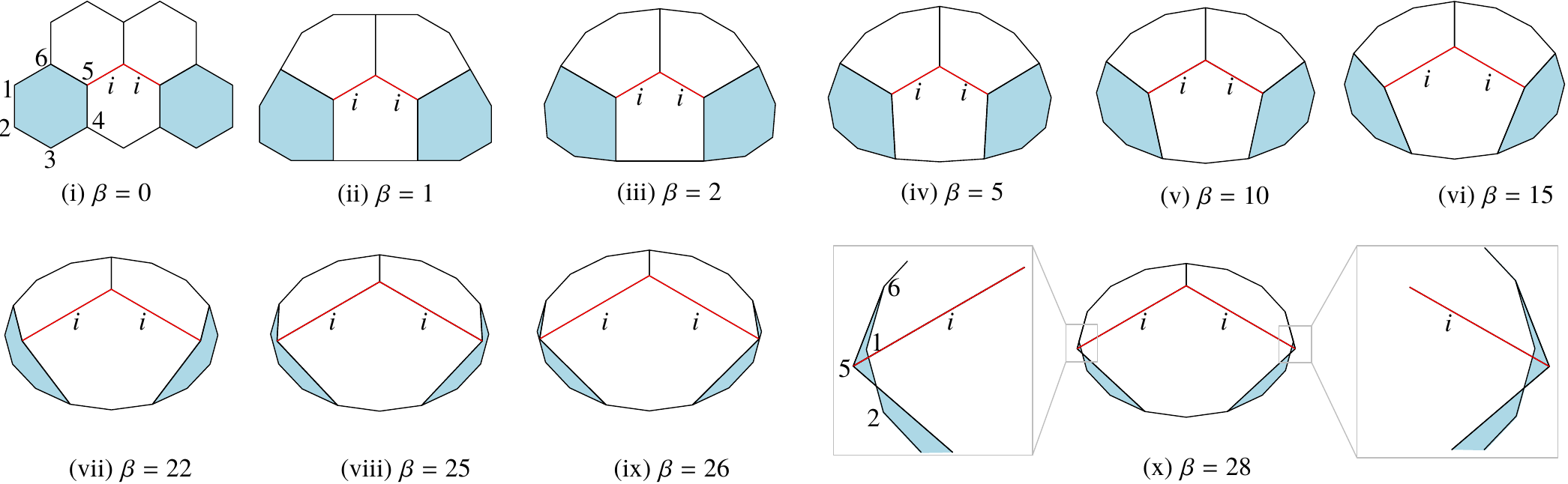}
		\caption{Higher values of $\beta$ can result in significant distortion of elements, and also, element flipping.}
		\label{fig:fig2d}
	\end{subfigure}
	
	\caption{V-notches furnish jumps in boundary normals ($\Gamma_\mathrm{{cm}}$) (Fig. \ref{fig:fig2a}) which are subdued via boundary smoothing (Fig. \ref{fig:fig2c}) to facilitate contact analysis. Fig. \ref{fig:fig2b} depicts the way boundary smoothing is performed. Fig. \ref{fig:fig2d} shows that for higher values of $\beta$, some elements may experience flipping.}
	\label{fig:continuityfigure}
\end{figure}

Systematic identification and shifting of boundary nodes is implemented within each iteration of the solution process. All boundary edges and nodes are first identified. Thereafter, mid-points of boundary edges are joined via straight lines. Boundary nodes are projected along their shortest perpendiculars (Fig. \ref{fig:fig2b}). This step can be performed multiple ($\beta $) times  where $\beta$ is specified prior to the analysis. Thus, new positions of the boundary nodes are obtained while retaining those for the interior nodes. Updated nodal coordinates are used in the finite element analysis. The elemental connectivity matrix is unaltered ensuring \textit{edge-connectivity} between cells. However, many cells get altered into concave elements. For the example shown (Fig. \ref{fig:fig2d}), at high values of $\beta$, internal edges (e.g., 4-5 and 5-6) first get straightened, and as $\beta$ is increased further, hexagonal cells at bottom left and right become concave in shape, and eventually flip at $\beta = 28$. Lengths of other internal edges (marked \textit{i}) do not change.  In our experience, $\beta = 10$ provides adequately smooth boundaries \cite{kumar2015topology}. For higher values, significant distortion or reduction in sizes of boundary elements may result. Also, some elements may flip, as shown in Fig. \ref{fig:fig2d}. Element flipping can be detected by noting that the element Jacobian becomes negative. Boundary smoothing may be executed until a stage where the Jacobian determinant becomes too low.  

Cells are removed in two steps: In the first, cells exposed to masks are removed and thereafter boundary smoothing is performed. As smoothing alters boundary cells but not the interior ones, the latter have original, regular shapes. In the second stage, these regular cells are also removed (Fig. \ref{fig:fig2b}). This is equivalent to placing additional negative masks over such cells. Removal in the second stage is done so that the constituting members become slender allowing them to undergo large deformation. An added advantage is the reduction in volume. As a consequence of this removal, new serrated boundaries get generated. At this stage, considering all remnant hexagonal cells in their original, regular forms, boundary smoothing is performed again, prior to further analysis.
Note that the removal of hexagonal cells is only temporary, to facilitate the contact analysis. Mean value coordinate shape functions \cite{floater2003mean,hormann2006mean} which can cater to polygonal elements of any shape, are employed in the finite element analysis \cite{sukumar2004conforming,sukumar2006recent}.

\section{FINITE ELEMENT FORMULATION WITH CONTACT}\label{CMFWWC}

\begin{figure}[h!]
	\centering
	\begin{subfigure}{0.45\columnwidth}
		\centering
		\includegraphics[scale=1.3]{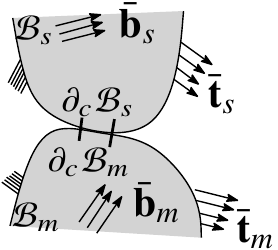}
		\caption{}
		\label{fig:fig_3a}
	\end{subfigure}
	\begin{subfigure} {0.45\columnwidth}
		\centering
		\includegraphics[scale=1.3]{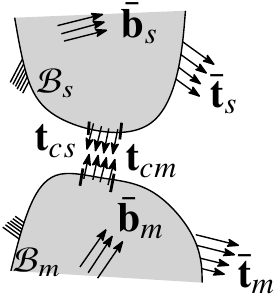}
		\caption{}
		\label{fig:fig_3b}
	\end{subfigure}
	\begin{subfigure}{0.45\columnwidth}
		\centering
		\includegraphics[scale=1.3]{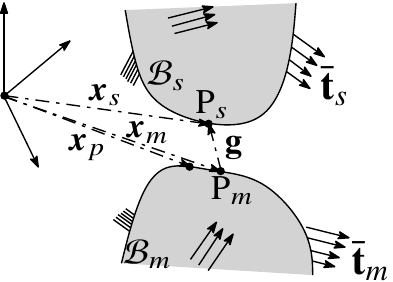}
		\caption{}
		\label{fig:fig_3c}
	\end{subfigure}
	\begin{subfigure}{0.45\columnwidth}
		\centering
		\includegraphics[scale=1.2]{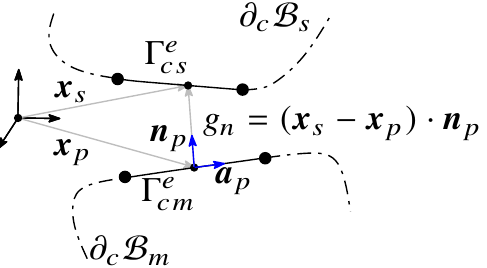}
		\caption{}
		\label{fig:fig_3d}
	\end{subfigure}
	\caption{A schematic diagram for the contact formulation. (\subref{fig:fig_3a}) Deformed configurations, (\subref{fig:fig_3b}) Contact description with tractions, (\subref{fig:fig_3c}) Closet point evaluation and (\subref{fig:fig_3d}) Contact discretization: piecewise linear segments approximate interacting boundaries}
	\label{fig:fig_4}
\end{figure}

The finite element formulation with contact is briefly reviewed. Consider two bodies $\mathcal{B}_k|_{k={s,m}}$ (current configurations) in contact, with  known sets of surface tractions $\bar{\bm{t}}_k$ on $\partial_t\mathcal{B}_k\subset \partial \mathcal{B}_k$, volumetric body loads $\bar{\bm{b}}_k$ in $\mathcal{B}_k$ and given deformation on $\partial_u\mathcal{B}_k\subset\partial\mathcal{B}_k$ (Fig. \ref{fig:fig_3a}).  The surface $\partial \mathcal{B}_k = \partial_u\mathcal{B}_k \cup \partial_t\mathcal{B}_k,\,\, \text{with}\,\,\,\partial_u\mathcal{B}_k \cap \partial_t\mathcal{B}_k = \emptyset$ where $\partial_u\mathcal{B}_k$ and $\partial_t\mathcal{B}_k$ are portions for prescribed displacements and tractions boundary respectively. Contact surfaces $\partial_c\mathcal{B}_k\subset\partial_t\mathcal{B}_k$ originate when bodies are in contact (Fig. \ref{fig:fig_3a}). 

The deformation field $\bm{u}_k \in \mathcal{U}_k$ is computed by satisfying the following weak form
\begin{equation}\label{eq:CMCMF1}
\displaystyle\sum_{k=1}^{2}\bigg[\delta\Pi_{\mathrm{int},\,k}+\delta\Pi_{\mathrm{c},\,k}-\delta\Pi_{\mathrm{ext},\,k}  \bigg]=0 \,\,\, \forall\delta\bm{ u}_k \in \mathcal{W}_k,
\end{equation}
\noindent where 
\begin{eqnarray}
\delta\Pi_{\mathrm{int},\,k} = \int_{\mathcal{B}_k} \bm{\sigma}_k: \text{grad}(\delta \bm{ u}_k)\mathrm{d} v_k\nonumber,\,\,
\delta\Pi_{\mathrm{c},\,k}= - \int_{\partial_c\mathcal{B}_k} \delta\bm{ u}_k \cdot \bm{t}_{ck}\mathrm{d}a_k, \nonumber \\
\mbox{and}\,\,\,\delta\Pi_{\mathrm{ext},\,k} =  \int_{\partial_t\mathcal{B}_k} \delta\bm{ u}_k \cdot \bar{\bm{t}}_{k}\mathrm{d}a_k + \int_{\mathcal{B}_k}\delta\bm{ u}_k\cdot\rho_k\bar{\bm{b}}_k \mathrm{d}v_k \nonumber 
\end{eqnarray} 

\noindent are the internal, contact and external virtual work, respectively. $\mathcal{U}_k$ and $\mathcal{W}_k$ are the sets of kinematically admissible deformations and variations, respectively. d$v_k$ and d$a_k$ denote elemental volumes and areas respectively, and $\bm{t}_{ck}$ is the contact traction on surface $\partial_\mathrm{c}\mathcal{B}_k$ arising due to contact (Fig. \ref{fig:fig_3b}). If one body is rigid, summation and index $k$ are dropped from Eq. \eqref{eq:CMCMF1}. $\bm{\sigma}_k$ represents the Cauchy stress tensor evaluated using the constitutive model of neo-Hookean material (strain energy function $W = \frac{\mu}{2}[\text{tr}\,(\bm{F}\bm{F}^T) -3 -2\ln J] + \frac{\Lambda}{2}(\ln J)^2$) \cite{zienkiewicz2005finite} as

\begin{equation}\label{eq:LDHB9}
\bm{\sigma} = \frac{\mu}{J}(\bm{F}\bm{F}^\mathrm{T}-\bm{I})+\frac{\Lambda}{J}(\ln J)\bm{I}
\end{equation}
where $\bm{F} = \text{Grad}\,\bm{u} + \bm{I}$ is the deformation gradient, $\mu = \frac{E}{2(1+\nu)}$ and $\Lambda = \frac{2\mu\nu}{1-2\nu}$ are Lame's constants, $J = \det(\bm{F})$, and $\bm{I}$ is the unit tensor. $\text{Grad}\,\bm{u}$ represents the gradient of $\bm{u}$ with respect to undeformed coordinates $\bm{X}$, and $E$ and $\nu$ are Young's modulus and Poisson's ratio, respectively. \\

The elemental displacement field $\bm{u}_e (=\bm{x}_e-\bm{X}_e)$ and corresponding variation $\delta\bm{u}_e$ is approximated\footnote{denoted via superscript $h$} as

\begin{equation}\label{eq:FEF1}
\begin{split}
\bm{u}_e \approx \bm{u}_e^h = \displaystyle\sum_{I=1}^{n_\mathrm{{nd}}=6} N_I \bm{u}_I = \mathbf{N}{\mathbf{u}}_e,\\ \delta\bm{u}_e \approx \delta\bm{u}_e^h = \displaystyle\sum_{I=1}^{n_\mathrm{{nd}}=6} N_I \bm{v}_I = \mathbf{N}{\mathbf{v}}_e,
\end{split}
\end{equation}  

\noindent Here $N_I$ are the mean value shape functions \cite{hormann2006mean} with $\mathbf{N} =\big[N_1\bm{I},\,\,N_2\bm{I},\,\,\cdots,N_{n_\mathrm{nd}}\bm{I}\big]$ and ${\mathbf{u}}_e =\big[\bm{u}_1^\mathrm{T},\,\,\bm{u}_2^\mathrm{T},\,\,\cdots,\bm{u}_{n_\mathrm{nd}}^\mathrm{T}\big]^\mathrm{T}$. $\bm{u}_I$ and $\bm{v}_I$ denote the nodal displacements and variations, respectively. Likewise, geometry of the element in undeformed ($\bm{X}_e$) and deformed ($\bm{x}_e$) configurations are approximated. In the discretized setting (Eq. \ref{eq:FEF1}), Eq. \eqref{eq:CMCMF1} yields

\begin{equation}\label{eq:FEF2}
\bm{v}^\mathrm{T}\big[\mathbf{f}_{\mathrm{int}}+ \mathbf{f}_{\mathrm{c}}-\mathbf{f}_{\mathrm{ext}}\big]  = 0 \quad\forall \bm{v} \in \mathcal{W}_k
\end{equation} 
where $\bm{v}$ is the global vector comprising of the kinematically admissible variation in nodal displacements $\bm{u}_I$. $ \mathbf{f}_{\mathrm{int}},\,\, \mathbf{f}_{\mathrm{c}},\,\,\mathbf{f}_{\mathrm{ext}}$ are internal, contact and external forces, respectively. Eq. \eqref{eq:FEF2} leads to nonlinear equilibrium equations $\mathbf{f}(\mathbf{u}) = \mathbf{f}_{\mathrm{int}}+ \mathbf{f}_{\mathrm{c}}-\mathbf{f}_{\mathrm{ext}} = \mathbf{0}$ ($\mathbf{f}(\mathbf{u})$ is the residual force) which are solved using the Newton-Raphson (N-R) iterative procedure. \\

The internal elemental forces $\mathbf{f}^e_{\mathrm{int}}$ are given as $ \int_{\mathcal{B}_k^e}\bm{B}_\mathrm{UL}^\mathrm{T}\bm{\sigma} \mathrm{d}\text{v},$
where $\mathbf{B}_\mathrm{UL}$ is the strain-displacement matrix \cite{wriggers2008nonlinear}. To evaluate the integral over each cell, the latter is divided into six triangular regions with respect to its centroid. 25 Gauss points are employed for integration over each triangular region for reasonable accuracy \cite{sukumar2004conforming} (see section \ref{CMPCSTs}). Evaluation of contact forces $\mathbf{f}_\mathrm{c}$ is described next. \\

\begin{table}[h!]
	\caption{Active constraint strategy for \textit{self} and \textit{mutual} contact} \label{T1}
	\begin{framed}
		Initialize: MutualContactPairsExist = 0;\, SelfContactPairsExist = 0; \\
		At each load step $r\rightarrow r+1$:\\
		Compute $\mathbf{f}_{ck}^e$ (Eq. \eqref{eq:CFCS11}) and $\mathbf{k}_{c,kl}^e|$ $(k=s,m),\,\, (l= m,s)$ for the contact surfaces $\Gamma_{ck}^e$ for each such quadrature point $\bm{x}_k \in\Gamma_{ck}^e$ (Fig. \ref{fig:fig_3d}):
		\begin{framed}
			\begin{enumerate}
				\item for closest projection point computation
				\begin{itemize}
					\item compute $\bm{x}_p = \bm{x}_m(\bm{\xi}_p)$ (Eq. \eqref{eq:CFCS8}).
					\item evaluate the normal gap $g_n$, normal vector $\bm{n}_p$ at $\bm{x}_p$ and corresponding $\bm{n}_p^s$ at $\bm{x}_s$. 
				\end{itemize}
				\item for contact identification
				\begin{itemize}
					\item In case of mutual contact \\
					if $\lambda \geq 0$,\, MutualContactPairsExist = 1, i.e.,
					new mutual contact pairs are detected
					\item In case of self contact \\
					if $\lambda > 0$\,\& $(\bm{n}_p^s\cdot \bm{n}_p)<0$,\,SelfContactPairsExist = 1 i.e., new self contact pairs are detected
				\end{itemize}
				\item for contact computation
				\begin{itemize}
					\item if contact constraint becomes active or new contact is detected, evaluate $\mathbf{f}_{ck}^e$ and $\mathbf{k}_{c,kl}^e$ and solve for equilibrium $\mathbf{f}(\mathbf{u})= \mathbf{0}$
					\item if contact constraint is inactive or previous contact is lost, $\mathbf{f}_{ck}^e$ and  $\mathbf{k}_{c,kl}^e$  are set to zero.
				\end{itemize}
			\end{enumerate}
		\end{framed}
	\end{framed}
\end{table}

Contact pairs are detected using the formulated \textit{active set strategy} of Table \ref{T1}. Contact tractions are treated as respective Lagrange multipliers via the augmented Lagrange multiplier method \cite{wriggers2006computational} combined with a segment-to-segment contact approach. Frictionless and adhesionless contact is considered. In augmented Lagrange multiplier method, contact tractions are modeled as

\begin{equation}\label{eq:CFCS3}
\bm{t}_{ck} =
\begin{cases}
\lambda \bm{n}_p  \qquad $\;\;$ g_n<0 \\
\mathbf{0} \qquad \qquad g_n\ge0,
\end{cases}
\end{equation}
where $\bm{n}_p$ denotes the normal vector at projection point $\bm{x}_p$ of point $\bm{x}_s\in \mathcal{B}_s$ and $\lambda = \lambda_{old} - \epsilon g_n$, is the Lagrange multiplier that is updated by an iteration. Here, $g_n = \mathbf{g}\cdot\bm{n}_p = (\bm{x}_s-\bm{x}_p)\cdot\bm{n}_p$ (Fig. \ref{fig:fig_3d}) is the normal gap. When evaluating $g_n$, one finds the closest projection point $\bm{x}_p$ on the  contact surface $\partial_\mathrm{c}\mathcal{B}_k|_{k={m,s}}$ corresponding to each point of the contact surface $\partial_\mathrm{c}\mathcal{B}_l|_{l={s,m}}$. 

For a known point $\bm{x}_s\,\in\partial_\mathrm{c}\mathcal{B}_s$, one searches for the closest projection point $\bm{x}_m\,\in \partial_\mathrm{c}\mathcal{B}_m$ (Fig. \ref{fig:fig_3c}). The search is performed by minimizing the distance $d = ||\bm{x}_s-\bm{x}_m||$ between points P$_s$ and P$_m$ in the convective coordinates setting $\bm{\xi} = (\xi^1,\,\xi^2)$ to represent each point on the surface $\partial_\mathrm{c}\mathcal{B}_m$ \cite{wriggers2006computational}. The tangent vectors at $\bm{x}_m$ are given as $\frac{\partial\bm{x}_m}{\partial\xi^\alpha}=\bm{a}_\alpha;\,\alpha = 1,\,2$. Minimization of $d$ gives two nonlinear equations

\begin{equation}\label{eq:CFCS8}
\frac{\partial d}{\partial{\xi}^\alpha} = -\big(\bm{x}_s-\bm{x}_m(\bm{\xi})\big)\cdot\bm{a}_\alpha = 0.
\end{equation}

\noindent Solving for these yields $\bm{\xi} =\bm{\xi}_p(=\xi_p^1,\xi_p^2)$ so that $\bm{x}_p=\bm{x}_m(\bm{\xi}_p)$ is the projection point closest to $\bm{x}_s$. The contact surface $\partial_\mathrm{c}\mathcal{B}_m$ at point  $\bm{x}_p$ is described via the co-variant tangent  vectors $\bm{a}_\alpha^p = \frac{\partial\bm{x}_p}{\partial\xi^\alpha}$  and the normal vector $\bm{n}_p = \frac{\bm{a}_1^p\times\bm{a}_2^p}{||\bm{a}_1^p\times\bm{a}_2^p||}$ \cite{wriggers2006computational}. Using $\bm{n}_p$, one can find the normal gap $g_n$ and contact tractions (Eq. \ref{eq:CFCS3}). In case of mutual contact, if $\lambda$ becomes positive, the contact constraint set to active, and thus contact forces are incorporated in the mechanical equilibrium conditions. In case $\lambda$ becomes negative, either previous contact is lost, or there is no contact. However, to detect self contact pairs, an additional check with normals is needed (section \ref{SCSA}). 

Contact surfaces $\partial_\mathrm{c}\mathcal{B}_k$ are approximated using surface (line) elements $\Gamma_{ck}^e$ containing two nodes at each such element (Fig. \ref{fig:fig_3d}). Displacements and corresponding variation fields are approximated with 2D linear Lagrange basis functions $N_I|_{I= 1,\,2}$ \cite{zienkiewicz2005finite}. In view of Eq. \eqref{eq:CMCMF1}, contact forces are evaluated using the full pass algorithm as
\begin{equation} \label{eq:CFCS11}
\mathbf{f}_{{ck}}^e = -\int_{\Gamma_{ck}^e} \mathbf{N}_k^T \bm{t}_{ck} \mathrm{d}a, \,\,\,\,\,\, \,\,\,\,\,\, \mathbf{f}_{{cl}}^e = \int_{\Gamma_{ck}^e} \mathbf{N}_l^T \bm{t}_{ck} \mathrm{d}a
\end{equation}
where $\Gamma_{ck}^e \subset \partial_c\mathcal{B}_k^h$, $\Gamma_{cl}^e \subset \partial_c\mathcal{B}_l^h$ and $\mathbf{N}_{k/l} = [N_{1}\bm{I}\quad N_{2}\bm{I}]$. Further, $N_{1} = \frac{1}{2}(1-\xi_{p})$ and $N_{2} = \frac{1}{2}(1+\xi_{p})$ with $\xi_{p}\in[-1\,\,1]$. If body $\mathcal{B}_m$ is rigid and motionless,  $\mathbf{f}_{cm}^e=\mathbf{0}$. Contact stiffness matrices $\mathbf{k}_{c,kl}^e|$ $(k=s,m),\,\, (l= m,s)$ are determined via linearization. A detailed derivation is given in \cite{kumar2017_diss}.
\subsection{Self contact search algorithm} \label{SCSA}
An active set strategy to detect both \textit{self} and \textit{mutual} contact surface pairs, within each Newton-Raphson iteration, is presented in Table \ref{T1}. In multi-body contact problems, mating surfaces (slave and master) are usually specified prior to the analysis making it relatively straightforward and efficient. In cases where members of a body themselves are deemed to come in contact, it is difficult to predetermine contact pairs. Identification of the latter depends on the topology of the body and loading boundary conditions. Such identifications become even more challenging with the evolution of topology of the body, as is the case with the synthesis of CCMs. It therefore becomes imperative to estimate contact pairs as part of the analysis, as opposed to specifying them a priori. 

\begin{figure}
	\centering
	\includegraphics[scale =1]{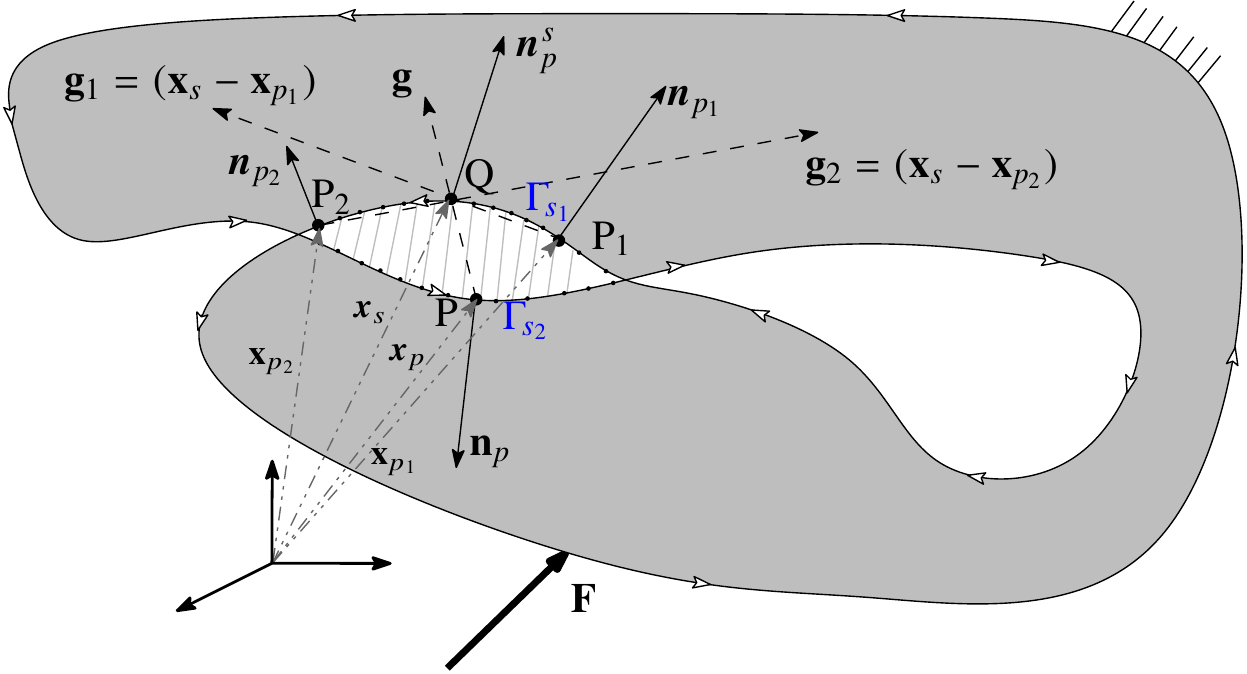}
	\caption{Schematic describing self contact when two surface regions $\Gamma_{s_{1}}$ and $\Gamma_{s_{2}}$  of the same body penetrate each other. The hatched portion shows that the top part of the body has penetrated into the bottom part. Arrows are marked on the boundary to indicate orientation and to differentiate between two interacting surfaces. For point $\bm{x}_{s}\in\Gamma_{s_{1}}$, point $\bm{x}_{p}\in\Gamma_{s_{2}}$ is the nearest neighbor. $\bm{x}_{p1}\in\Gamma_{s_{1}}$ or $\mathrm{x}_{p2}\in\Gamma_{s_{1}}$ are also candidate nearest neighbors since normal gaps $\mathbf{g}_1\cdot\bm{n}_{p_{1}}$ and $\mathbf{g}_2\cdot\bm{n}_{p_{2}}$ are negative in both cases. Such candidate nearest neighbors are discarded noting that the dot products between normals, e.g., $\bm{n}_\mathrm{{p}}^s\cdot\bm{n}_{p_{1}}$ and $\bm{n}_\mathrm{{p}}^s\cdot\bm{n}_{p_{2}}$ are positive.} 
	\label{fig:4}
\end{figure}

To determine self contact pairs, efficient algorithms of order $O(nlogn)$ exist, e.g., in [42], though an $O(n^2)$ strategy, termed as global or nearest neighboring search \cite{knuth1998art}, is presented here, notwithstanding efficiency. For any point $\bm{x}_{s}$ on the boundary (interior or exterior) of the body as reference, its nearest neighbor, $\bm{x}_{p}$, not the same as $\bm{x}_{s}$, is determined by solving Eq. \ref{eq:CFCS8}. The tangent vectors are evaluated at $\bm{x}_{p}$, and the normals at both, $\bm{x}_{s}$ and $\bm{x}_{p}$. The normal gap between $\bm{x}_{s}$ and $\bm{x}_{p}$ is computed. The dot product between normals at $\bm{x}_{s}$ and $\bm{x}_{p}$ is also evaluated. If both, the normal gap and dot product are negative, boundary elements (line segments) containing $\bm{x}_{s}$ and $\bm{x}_{p}$ intersect. In that case, the two elements constitute a self contact pair for which contact forces and stiffness matrices are computed. If the normal gap is negative but the dot product is positive, (e.g., for points $\bm{x}_{p1}$ and $\bm{x}_{p2}$ in Fig. \ref{fig:4}), corresponding boundary elements do not intersect. Identification of self contact line pairs is performed for all points on the continuum boundary(ies).

\section{SYNTHESIS OF CONTACT-AIDED COMPLIANT CONTINUA}\label{PF}
\begin{figure*}
	\centering
	\includegraphics[scale=1]{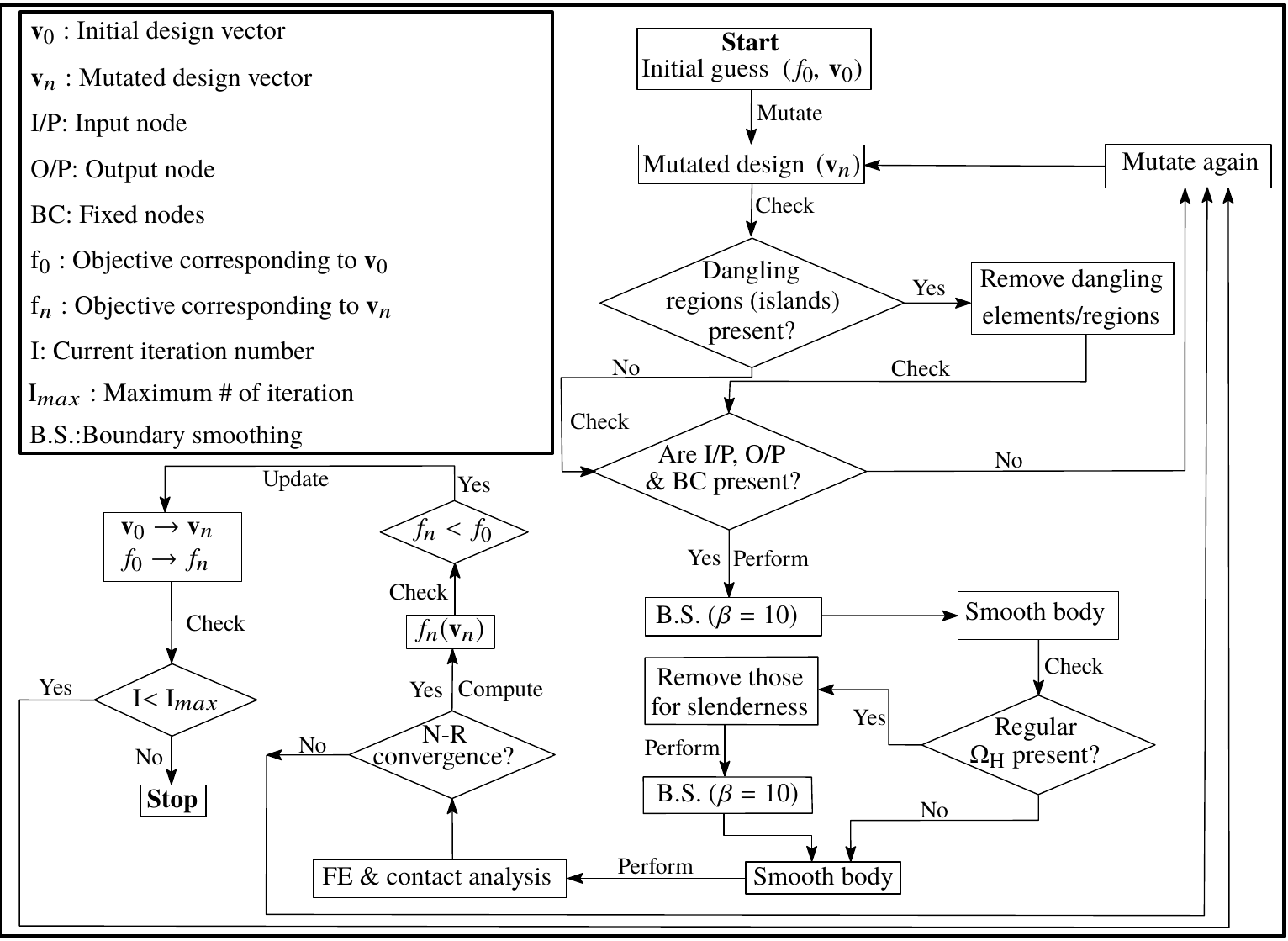}
	\caption{A schematic flowchart for a single design iteration within the proposed optimization approach}
	\label{fig:5}
\end{figure*}

The design approach is illustrated via synthesis of four large, $C^0$ path generating CCMs. Capability of the approach is demonstrated by synthesizing a CCM that can trace a $Z-$path (multiple kinks), exemplifying the possibility of obtaining intricate deformation characteristics. CCMs are obtained by minimizing the FSDs objective \cite{zahn1972fourier} using a stochastic hill climber search \cite{Russell:2003:AIM:773294}. The design variables used are the mask parameters (section \ref{NCMnew}). The magnitude of the input force (along with the possibility of force reversal along a prescribed direction) is also considered a design variable. The overall schematic is depicted via a flow chart in Fig. \ref{fig:5}. 

\subsection{Fourier Shape Descriptors (FSDs) objective} \label{FSDobj}
Traditionally, an objective based on the sum of the squares of the difference between  coordinates of the constitutive precision points of the desired and actual paths is used to synthesized path generating mechanisms. Though  simple to implement, it has shortcomings \cite{ullah1997optimal,mankame2007synthesis}. It suffers from timing constraints and does not permit individual control on shape, size and initial orientation of the path. The FSDs objective \cite{ullah1997optimal} offers more flexibility  in that shapes of the two paths can be compared independent of the number of precision points used to specify them. Shape and size measures can also be decoupled. This objective is minimized herein to synthesize CCMs. An FSDs objective is evaluated from the Fourier coefficients of the specified and actual paths. To determine the respective coefficients, the paths are closed in clockwise sense such that they do not self-intersect, and parameterized as functions of the respective normalized arc length parameters \cite{zahn1972fourier}. \\

Let $A^j_m$ and $B^j_m$ be the Fourier coefficients, $L^j$ be the total length and $\theta^j$ be the initial orientation of two paths, $j=s\, \text{(specified)},\,a\, \text{(actual)}$. The objective used is

\begin{equation}\label{Eq:objective}
f(\mathbf{v}) = w_a A_\mathrm{{err}}+ w_b B_\mathrm{{err}}+ w_L L_\mathrm{{err}}+ w_\theta \theta_\mathrm{{err}}
\end{equation}

\noindent where $w_a,\,w_b,\,w_L,\,\text{and}\,w_\theta$ are user defined weights (Table \ref{T2}) for errors $A_\mathrm{{err}},\,B_\mathrm{{err}},\,L_\mathrm{{err}}\,\,\text{and}\,\,\theta_\mathrm{{err}}$, respectively (Eq. \ref{eq:error}). $A_\mathrm{{err}}\,\,\text{and}\,\,B_\mathrm{{err}}$ are the errors in the Fourier coefficients \cite{rai2007synthesis} that quantify discrepancy in shape. The last two error measures capture the difference in length (size) and initial orientation of the desired path. Larger weights are used to capture path shapes, while relatively smaller weights are used for path lengths and much smaller weights are used for path orientation, with the reasoning that a CCM can be rotated in order for the orientation of the desired path and that achieved to be alligned. Inverse problems, such as those posed by Eq. (\ref{Eq:objective}) can yield numerous solutions \cite{tikhonov2013numerical}. With multiple possibilities for contact, both in type (\textit{self} and \textit{mutual}) and number, a single set of weights for identical specifications can yield multiple solutions for CCMs (e.g., Figs \ref{fig:7} (c) and \ref{fig:fig_12a}) in that the corresponding design space is expected to be non-convex. 

\noindent The errors are defined as

\begin{equation}\label{eq:error}
\begin{aligned}
A_\mathrm{{err}}&=\sum_{m=1}^{N}(A_m^s-A_m^a)^2,\qquad
B_\mathrm{{err}}=\sum_{m=1}^{N}(B_m^s-B_m^a)^2,\\\
L_\mathrm{{err}}&=(L^s-L^a)^2,\qquad \quad
\theta_\mathrm{{err}}=(\theta^s-\theta^a)^2.
\end{aligned}
\end{equation}
Here,  $N = 50$ is the number of Fourier coefficients used. The optimization problem using the FSDs objective is stated as:

\begin{equation}\label{Eq:optimization}
\begin{aligned}
& \underset{\mathbf{v}}{\text{min}}
& &f(\mathbf{v}) + \lambda_v (V-V^*),\\
& \text{such that} & & \mathbf{f}(\mathbf{u}) = \mathbf{0};\,\, z_L\le z_i\le z_U|_{z_i = x_i,\,y_i,\,r_i}\\ 
&\,& &s_i\,(= 0\,\, \text{or}\,\, 1)\,;\,\, f_i\,[\in (0,1)]
\end{aligned}
\end{equation}

where $\mathbf{v}$ is the design vector, and $V$ and $V^*$ are the current and permitted volumes of the continua, respectively. In general, volume penalization parameter $\lambda_v$ should be taken relative to the objective. For the examples herein, it is taken as 20 if $V\ge V^*$, otherwise $\lambda_v = 0$. $\mathbf{f}(\mathbf{u}) = \mathbf{0}$ is the mechanical equilibrium equation (Eq. \ref{eq:FEF2}). $z_L$ and $z_U$ are the lower and upper bounds for $z_i \in \mathbf{v}$ (section \ref{NCMnew}). $M$ is the total number of overlaying masks kept constant throughout the optimization (see section \ref{MSKADDDEL}).

\subsection{Hill-climber search}\label{HC}
Evolution of the CCM continua is performed via a random mutation based stochastic, hill-climber search \cite{Russell:2003:AIM:773294}. However, in general, one can use any stochastic search algorithm. The cardinal reason is to evaluate a realizable, smooth, perfectly binary design in each search iteration using  large deformation nonlinear contact finite element analysis.  For $M$ negative circular masks, a design vector $\mathbf{v}$ contains $5M$ variables (section \ref{TOCSG}). Positions (center coordinates) and sizes (radii) of these masks define a potential CCM in each optimization iteration, and $s_i,\,f_i$ from the $i^\mathrm{th}$ mask provide status and size of contact surface within that mask. To mutate the design vector, one sets a small probability mutation number $pr \; (= 8\%)$. A random number $\eta$ is generated for each variable $p$. If $\eta<pr$, the variable is mutated as $p = p \pm (c\times m_{\mathrm{max}})$; otherwise $p$ remains unaltered. Here, $c$ is a random number and $m_{\mathrm{max}}$ is set to $15\%$ of the \textit{max}$(L_x,\,L_y)$. $L_x\,\text{and}\,L_y$ represent the design domain dimension in horizontal and vertical directions, respectively. For $s_i$, if $\eta<pr$ and $c>0.5$, $s_i = 1$, otherwise $s_i = 0$. $f_i$ is mutated similarly in $[0, 1)$. After mutation, one gets the new design vector $\mathbf{v}_n$. The FSDs objective $f(\mathbf{v}_\mathrm{new})$ is evaluated if the corresponding new CCM is well connected, has all input and output ports and requisite number of fixed degrees of freedom, and if the contact analysis converges. Else, the degenerate CCM solution is penalized and a new solution is sought. If $f(\mathbf{v}_\mathrm{new})<f(\mathbf{v})$ then $\mathbf{v}\leftarrow \mathbf{v}_\mathrm{new}$. Magnitude F  of the input force is
also taken as a design variable \cite{mankame2007synthesis,kumar2015synthesis}  and is mutated as $F = F \pm (c\times m_{\mathrm{max}})$. The input force is only permitted to flip along the prescribed direction.


\subsection{Evolved CCMs} \label{SE}
\begin{figure}
	\centering
	\begin{subfigure}{0.45\columnwidth}
		\centering
		\includegraphics[scale=0.7]{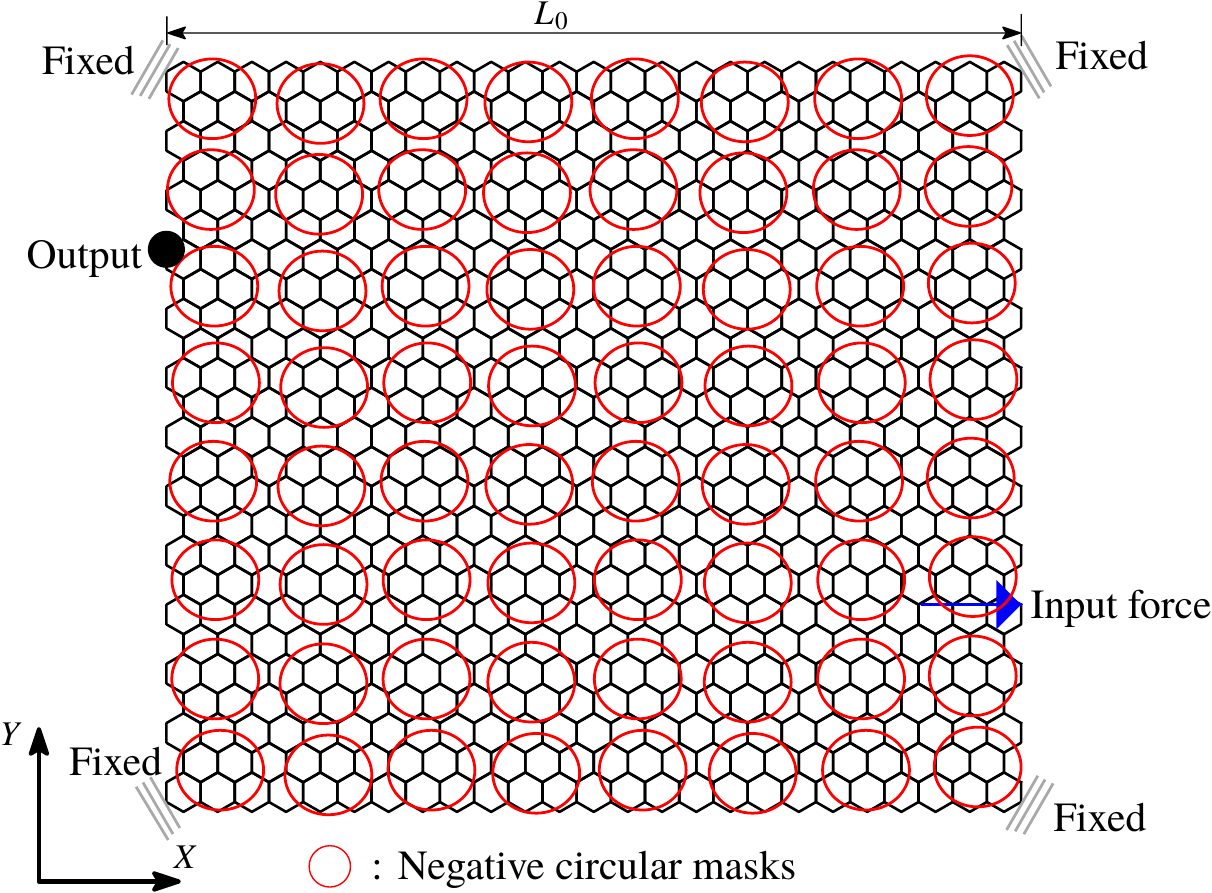}
		\caption{Design specification for all examples with the initial guess for mask parameters}
		\label{fig:6a}
	\end{subfigure}
	\begin{subfigure}{0.45\columnwidth}
		\centering
		\includegraphics[scale=0.6]{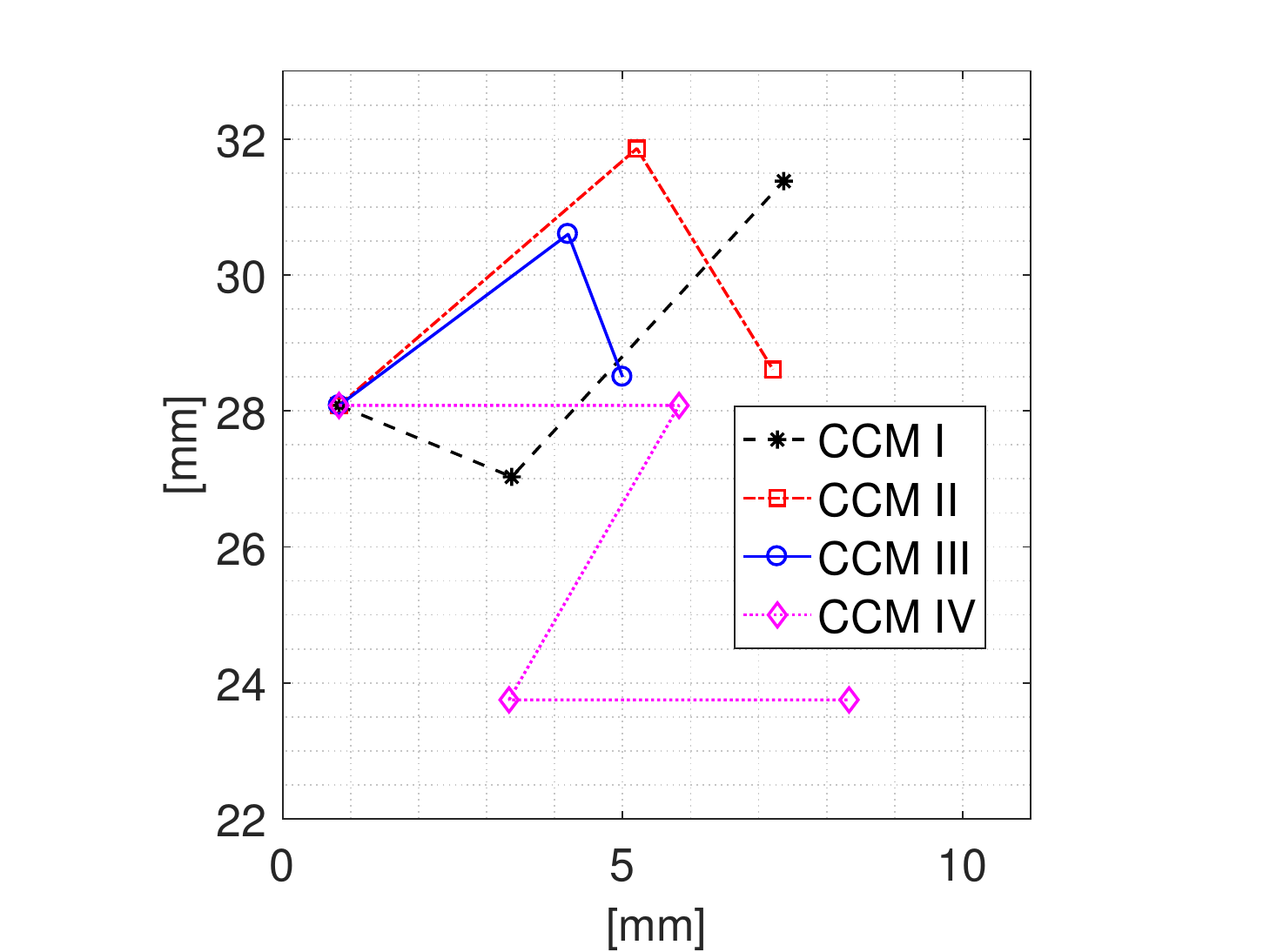}
		\caption{Desired output paths for all four CCMs are depicted}
		\label{fig:6b}
	\end{subfigure} 
	\caption{Design specification and specified paths for four different CCMs}
	\label{fig:6}
\end{figure}

Four examples of path generating Contact-aided Compliant Mechanisms are presented using identical design specifications but with different, long ($>10\%$ of the characteristic length, $L_0$), specified non-smooth paths to demonstrate the versatility of the synthesis approach. The specified paths are shown in Fig. \ref{fig:6b}.  The design requirements are depicted in Fig. \ref{fig:6a} and the associated parameters are depicted in Table \ref{T2}. We initiate the optimization with an actuating input force of $100$ N in the positive horizontal direction.

\begin{table}
	\centering
	\caption{Parameters used for synthesizing CCMs.}
	\label{T2}
	\resizebox{\columnwidth}{!}{%
		\begin{tabular}{  l      c       c  } 
			\hline
			\textbf{Parameter's name} & \textbf{Units} & \textbf{Value}  \\ \hline
			Design space $(\Omega)$  &$-$  & $25\Omega_\mathrm{H} \times25\Omega_\mathrm{H} $  \\ 
			number of masks in horizontal direction ($\mathrm{N}_\mathrm{x}$) & $-$ & $8$  \\ 
			number of masks in vertical direction ($\mathrm{N}_\mathrm{y}$) & $-$& $8$ \\
			Maximum radius of masks & mm& $8.0$\\
			Minimum radius of masks & mm& $0.1$\\
			Maximum number of function evaluations &$-$& $20000$ \\ 
			Young's modulus ($E_0$) & MPa & $2100$ \\ 
			Poisson's ratio &$-$ & $0.33$ \\ 
			Permitted volume fraction ($\frac{V^*}{V}$)&$-$ & $0.30$\\
			Mutation probability($pr$) &$-$& $0.08$ \\ 
			Contact surface radii factor (max($f_i$)) &$-$& $0.90$ \\ 
			Maximum mutation size ($m_\mathrm{{max}}$) &$-$& $6$\\ 
			Upper limit of the input load ($\bm{\mathrm{F}}_U$) & N&$500$\\ 
			Lower limit of the input load ($\bm{\mathrm{F}}_L$) & N& $-500$\\ 
			Weight of $a_\mathrm{err}$ ($w_{a}$) & rad$^{-2}$& $100$ \\ 
			Weight of $b_\mathrm{err}$ ($w_{b}$) & rad$^{-2}$&$100$\\ 
			Weight of path length error ($w_{L}$) & mm$^{-2}$& $1$\\ 
			Weight of path orientation error ($w_{\theta}$) &rad$^{-2}$& $0.1$\\ 
			number of Fourier coefficients&$-$ &$50$\\
			Boundary smoothing steps ($\beta$)& $-$& $10$ \\ 
			Maximum characteristic length (max$(L_x,\,L_y)$)& mm& $25\sqrt{3}$ \\ 
			Penalty parameter, mutual contact ($\epsilon_n$)& Nmm$^{-3}$& $50{E_0}/{L_0}$ \\ 
			Penalty parameter, self contact ($\epsilon_s$)& Nmm$^{-3}$& $4{E_0}/{L_0}$ \\ 
			\hline
		\end{tabular} 
	}
\end{table}  

\subsection{CCM Continua} \label{CCM_continua}
\begin{figure*}
	\centering
	\includegraphics[scale=1]{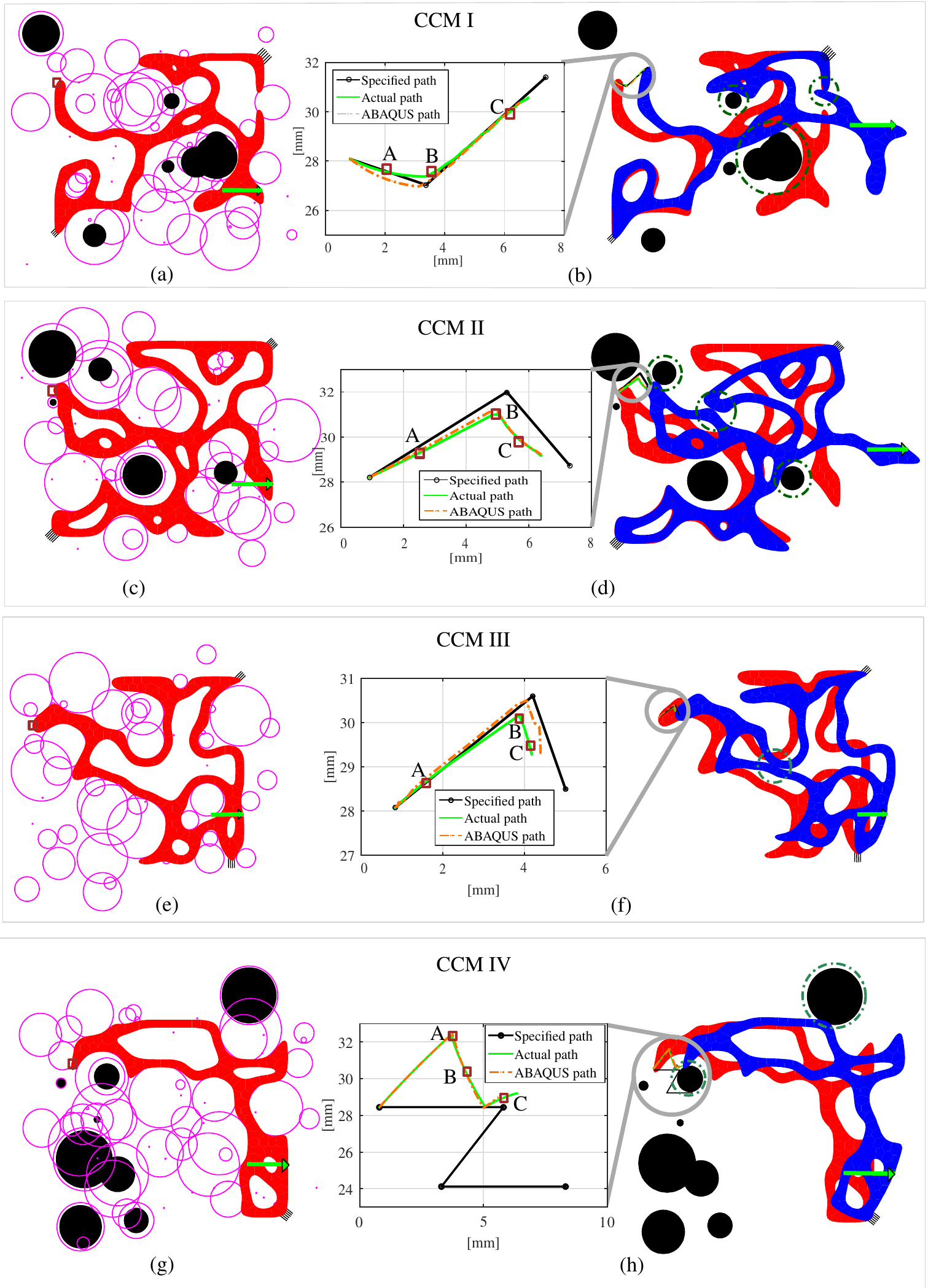}
	\caption{Final solutions for CCM I-IV: Left Column: Final continua with positions of masks and mutual contact surfaces shown, Middle and right columns: Comparison of paths generated and final deformed configurations. Active contact regions are depicted within dashed circles.}
	\label{fig:7}
\end{figure*}

\begin{figure}
	\centering
	\begin{subfigure}{0.45\columnwidth}
		\centering
		\includegraphics[scale=0.45]{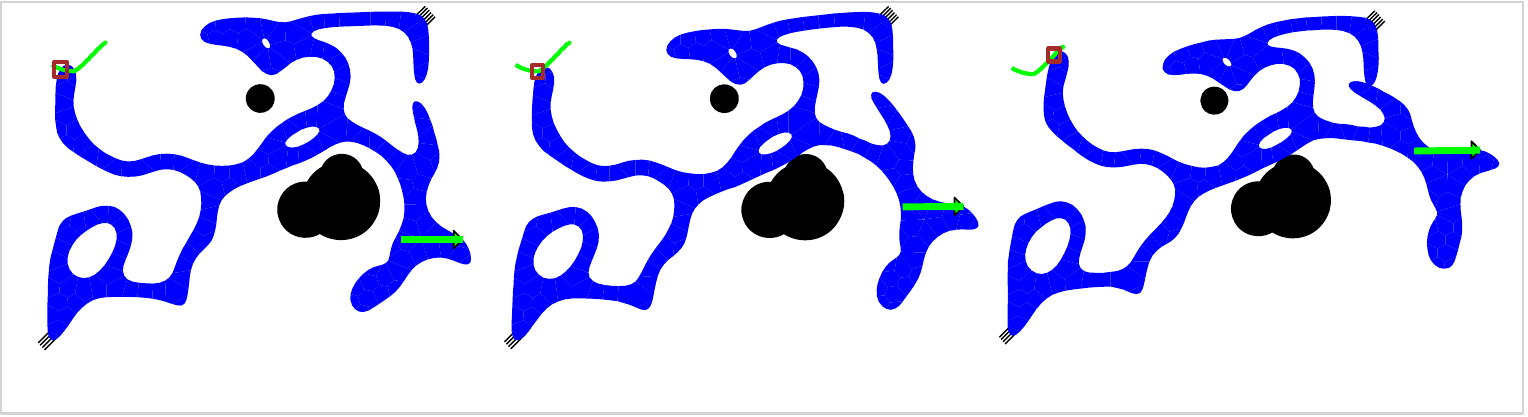}
		\caption{CCM I}
		\label{fig:8a}
	\end{subfigure}
	\begin{subfigure}{0.45\columnwidth}
		\centering
		\includegraphics[scale=0.45]{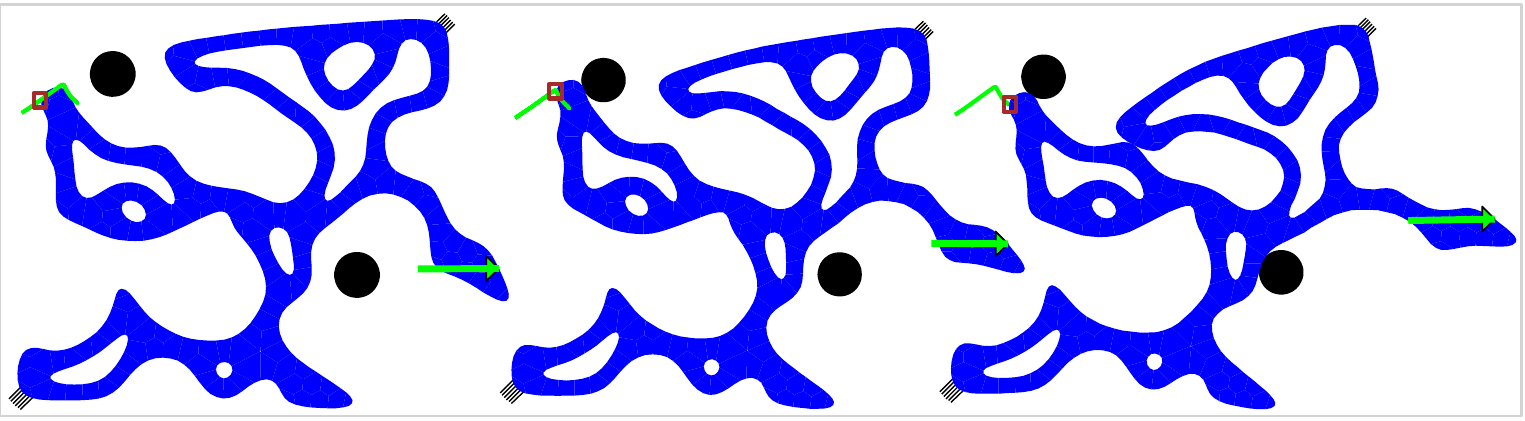}
		\caption{CCM II}
		\label{fig:8b}
	\end{subfigure}
	\begin{subfigure}{0.45\columnwidth}
		\centering
		\includegraphics[scale=0.45]{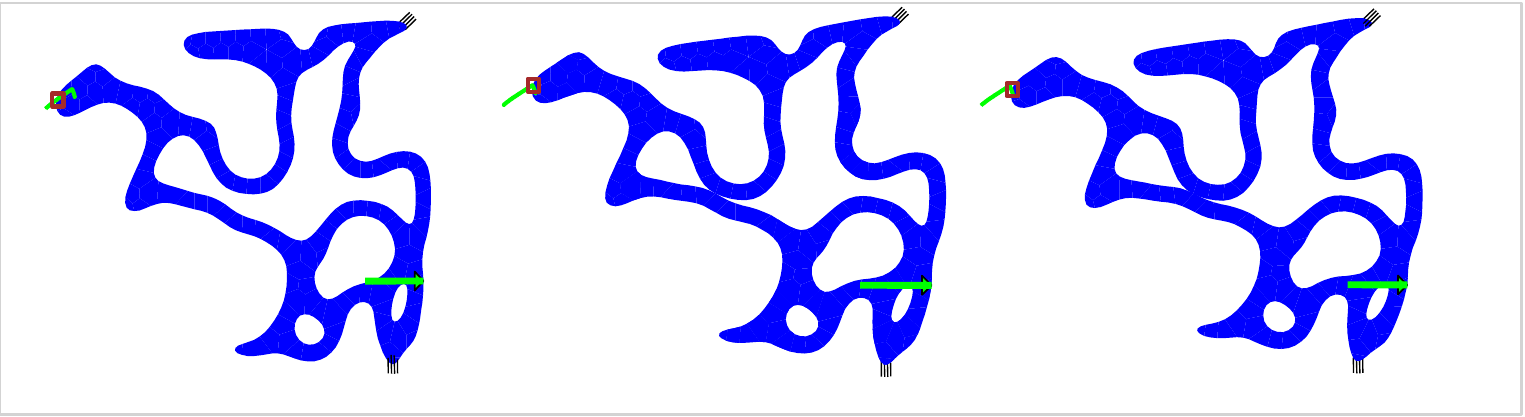}
		\caption{CCM III}
		\label{fig:8c}
	\end{subfigure}
	\begin{subfigure}{0.45\columnwidth}
		\centering
		\includegraphics[scale=0.45]{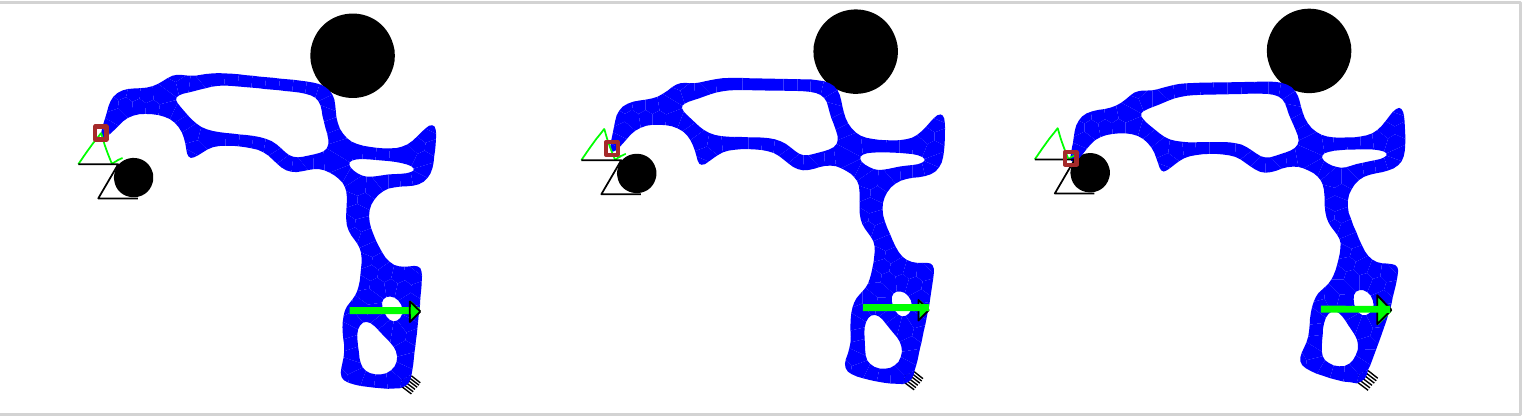}
		\caption{CCM IV}
		\label{fig:8d}
	\end{subfigure}
	\caption{Deformed configurations of CCMs at stages A, B and C (Fig. \ref{fig:7})}
	\label{fig:8}
\end{figure}

The final solutions of the four CCMs for the specified paths in Fig. \ref{fig:6b} are shown  in Fig. \ref{fig:7}a.  Optimal topologies of CCMs and negative masks are depicted. To generate CCM III, $7 \times 7$  masks are used and for other CCMs, $8\times8$ masks are employed. Masks suspending rigid contact surfaces (black, filled  circles) are also depicted. Actuating forces and (remnant) fixed  boundaries are annotated. Note that not all fixed boundaries specified prior to optimization are retained in the final designs.

CCM I is obtained after $4375$ search iterations with $131.05$ N input force in the positive horizontal direction. CCMs II,  III and IV are achieved after $3891$, $6673$ and $8675$ search iterations with the required input forces $134.65$ N, $97.91$ N  and $189.60$ N along the positive horizontal direction, respectively. The undeformed  and deformed (blue) configurations are shown with active contact locations encircled within the dash-dotted gray circles (Fig. \ref{fig:7}). 

To compare errors in shape between the specified and actual paths of the CCMs, we use $A_\mathrm{err}$ and $B_\mathrm{err}$ (Eq. \ref{eq:error}).  In addition, $\zeta_{l}$ and $\theta_\mathrm{diff} = |\theta^s - \theta^a|$ are used to show deviation in lengths and initial orientations between both paths. 
Here, superscripts $s\,\text{and}\, a$ represent the specified and actual paths, respectively and $\zeta_l$ is defined as:
\begin{equation} \zeta_{l} = \frac{|L^{s} - L^{a}|}{L^{s}} \times 100 \%.
\end{equation} 

\begin{table}
	\centering
	\caption{Errors in FSDs coefficients, lengths and orientations of the paths traced by CCMs with respect to their corresponding desired paths}
	\label{T3}
	\begin{tabular}{  l c  c c c} 
		\hline
		\textbf{Mechanisms} &  $A_\mathrm{err}$ & $B_\mathrm{err}$&\textbf{$\zeta_l$} (\%) &$\theta_\mathrm{diff}$(degree)  \\ \hline
		CCM I  & 0.0613 & 0.0750 &13.5282 & 6.2848\\ 
		CCM II  & 0.1879 & 0.1559 &19.6902& 7.2912\\ 
		CCM III  & 0.0332 & 0.5312& 23.3727 & 4.9178\\
		CCM IV & 0.5486 & 0.5581& 30.1120 & 56.1630\\
		\hline
	\end{tabular} 
\end{table}  
Notwithstanding orientation and size (length), in most cases, the actual paths compare well (Table \ref{T3}) with the respective desired paths in shape. In all cases, discrepancies in length ($\zeta_l$) of the paths are large, possibly because the weight $w_L$ for $L_{\mathrm{err}}$ used is significantly smaller ($w_L = 1$) than that ($w_a = w_b = 100$) used for the FSD coefficients (Table \ref{T2}). CCM IV traces a $Z-$path, representative of how intricate the deformation characteristics can be achieved by the CCMs designed using the proposed approach. While the overall shapes agree well (Table \ref{T3}), discrepancy is primarily due to path size (the segment after the second kink is smaller in size than desired) and the initial orientation ($\theta_\mathrm{diff} = 56.1630^o$) of the actual $Z$-path is not the same as that of the desired one (Fig. \ref{fig:7}h). Reasons for these differences could be the lower weights $(w_l = 1, w_\theta = 0.1)$ used, as, only capturing shape and size of the desired path is intended primarily. For CCMs I, II and III, discrepancies in initial orientations with respect to the respective desired paths are within $8\%$.

The length of the actual path obtained via the presented approach is about $7.40$ mm for CCM I, $7.30$ mm for CCM II, $4.75$ mm for CCM III, and  $10.50$ mm for CCM IV. These correspond to approximately $17\%,\, 16.9\%,\,11\%$  and $24\%$ of the maximum characteristic size ($25\sqrt{3}$ mm) of the design regions suggesting that all presented CCMs undergo large deformation. In addition, mechanisms also experience \textit{self} and \textit{mutual} contacts at various time steps in their deformation histories. Deformation profiles of these CCMs at stages A, B and C (Fig. \ref{fig:7}--middle column)  are depicted in Fig. \ref{fig:8}. The synthesis algorithm suggests many rigid contact surfaces, however, only some participate in active contact (Fig. \ref{fig:7}--third column). CCM I, when deforming, first experiences mutual contact (Fig. \ref{fig:8a}).  Thereafter, two of its branches interact with each other  which  is followed by mutual contact with another rigid surface (Fig. \ref{fig:7}b). CCM II first comes in contact with two rigid contact surfaces at different time instances (Fig. \ref{fig:8b}) and thereafter two of its constituting members interact with  each other (Fig. \ref{fig:7}d). With CCM III, self contact is observed with the potential of another self contact site on top left (Fig. \ref{fig:7}f). CCM IV experiences large  deformation with two mutual contacts at different temporal configurations, both, mandatory to yield the two desired kinks (Fig. \ref{fig:7}h). 

\begin{figure}
	\centering
	\begin{subfigure}{0.95\columnwidth}
		\centering
		\includegraphics[scale=1]{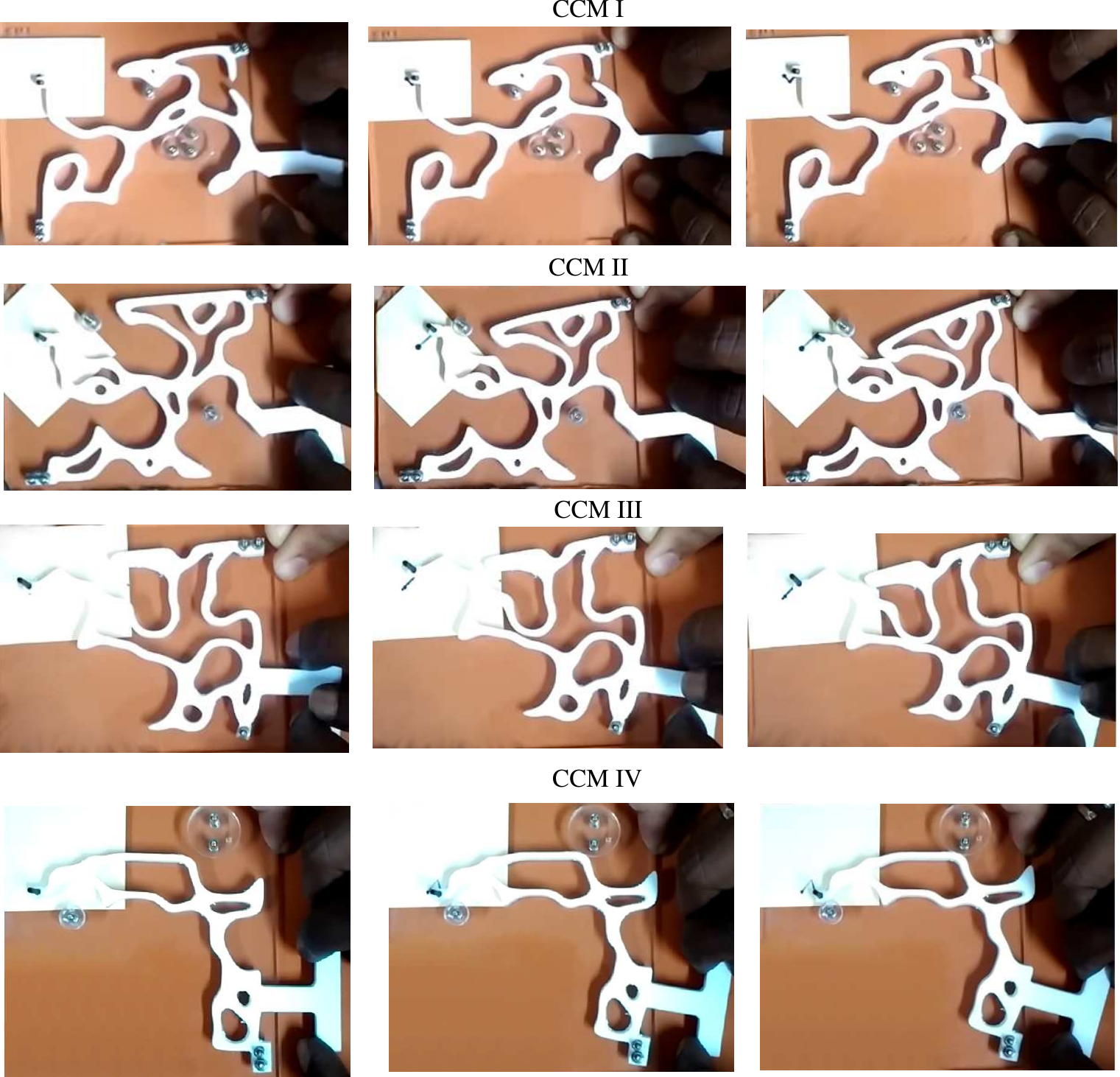}
		\caption{ }
		\label{fig:9a}	
	\end{subfigure}
	\begin{subfigure}{0.95\columnwidth}
		\includegraphics[scale=0.75]{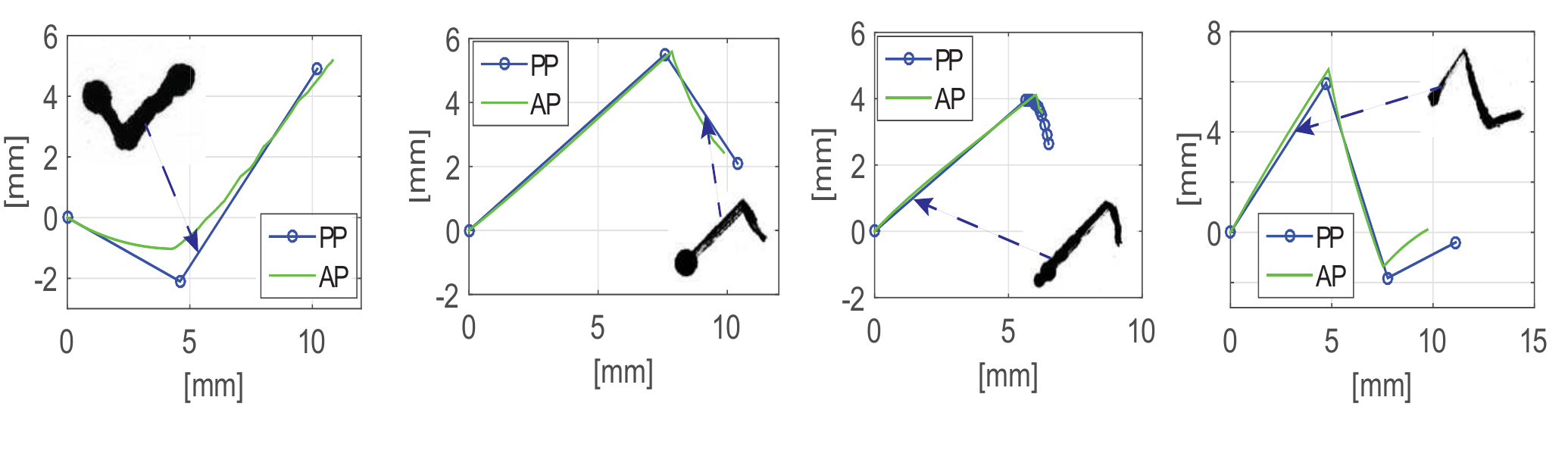}
		\caption{ }
		\label{fig:9b}
	\end{subfigure}
	\caption{(a) Prototypes of CCMs I-IV in their undeformed (left) and deformed (middle and right) configurations (b) Paths from the prototypes (PP) compared with those traced by the CCMs (AP) in simulation. left-right: CCMs I-IV.}
	\label{fig:9}
\end{figure}

Prototypes of these  CCMs (scale 2:1) are manufactured and tested. Their undeformed and deformed configurations are depicted in Fig \ref{fig:9}. Low values of $A_\mathrm{err},\,B_\mathrm{err}$ and $\zeta_l$ (Table \ref{T4}) indicate good agreement in shapes and sizes of the paths traced.  We also perform nonlinear contact finite element analysis using ABAQUS. The paths obtained are overlaid with the actual and desired paths and depicted in Fig. \ref{fig:7} (middle column) and the difference in the FSD coefficients and the length comparison are reported in Table \ref{T5}. One notices that $\,B_\mathrm{err}$ and the length error $\zeta_l$ for CCM III are relatively high (section \ref{coarse_fine_mesh}). Possible reason for high $B_\mathrm{err}$ is an additional kink observed in the path when CCM III is analyzed using ABAQUS (Fig. \ref{fig:7}f), and higher $\zeta_l$ is due to difference in path lengths.

\begin{table}
	\centering
	\caption{Errors in FSD coefficients, lengths and orientations of the paths traced by the manufactured prototypes and respect to their corresponding models}
	\label{T4}
	\begin{tabular}{l c  c c c} 
		\hline
		\textbf{Mechanisms} &  $A_\mathrm{err}$ & $B_\mathrm{err}$&\textbf{$\zeta_l$} (\%) &$\theta_\mathrm{diff}$(degree)  \\ \hline
		CCM I  & 0.1572 &0.2114 & 0.8754& 0.0938\\ 
		CCM II  & 0.0578 & 0.0637 &3.3197& 3.0436\\ 
		CCM III  & 0.0207 & 0.1866& 3.3776 & 9.9943\\
		CCM IV & 0.1950 & 0.1933& 2.8241 & 3.6853\\
		\hline
	\end{tabular} 
\end{table}  

\begin{table}
	\centering
	\caption{Errors in FSD coefficients, lengths and orientations of the paths traced via CCMs and that obtained using ABAQUS}
	\label{T5}
	\begin{tabular}{  l c  c c c} 
		\hline
		\textbf{Mechanisms} &  $A_\mathrm{err}$ & $B_\mathrm{err}$&\textbf{$\zeta_l$} (\%) &$\theta_\mathrm{diff}$(degree)  \\ \hline
		CCM I  & 0.0883 &0.1250 &  2.2428& 14.3154\\ 
		CCM II  & 0.0366& 0.0347 &2.3377& 1.2158\\ 
		CCM III  & 0.0579& 6.7219& 10.7810 & 4.174\\
		CCM IV & 0.0609 & 0.0647& 1.5912 & 0.5231\\
		\hline
	\end{tabular} 
\end{table}  

\section{DISCUSSION}\label{con}

This paper presents a continuum based synthesis approach for geometrically/materially large deformation CCMs that incorporate \textit{self} and \textit{mutual} contact modes. We discuss below various aspects of the approach.

\subsection{Performance with finer meshes}
\label{coarse_fine_mesh}
\begin{figure}[h!]
	\begin{subfigure}{0.45\textwidth}
		\centering
		\includegraphics[scale = 0.5]{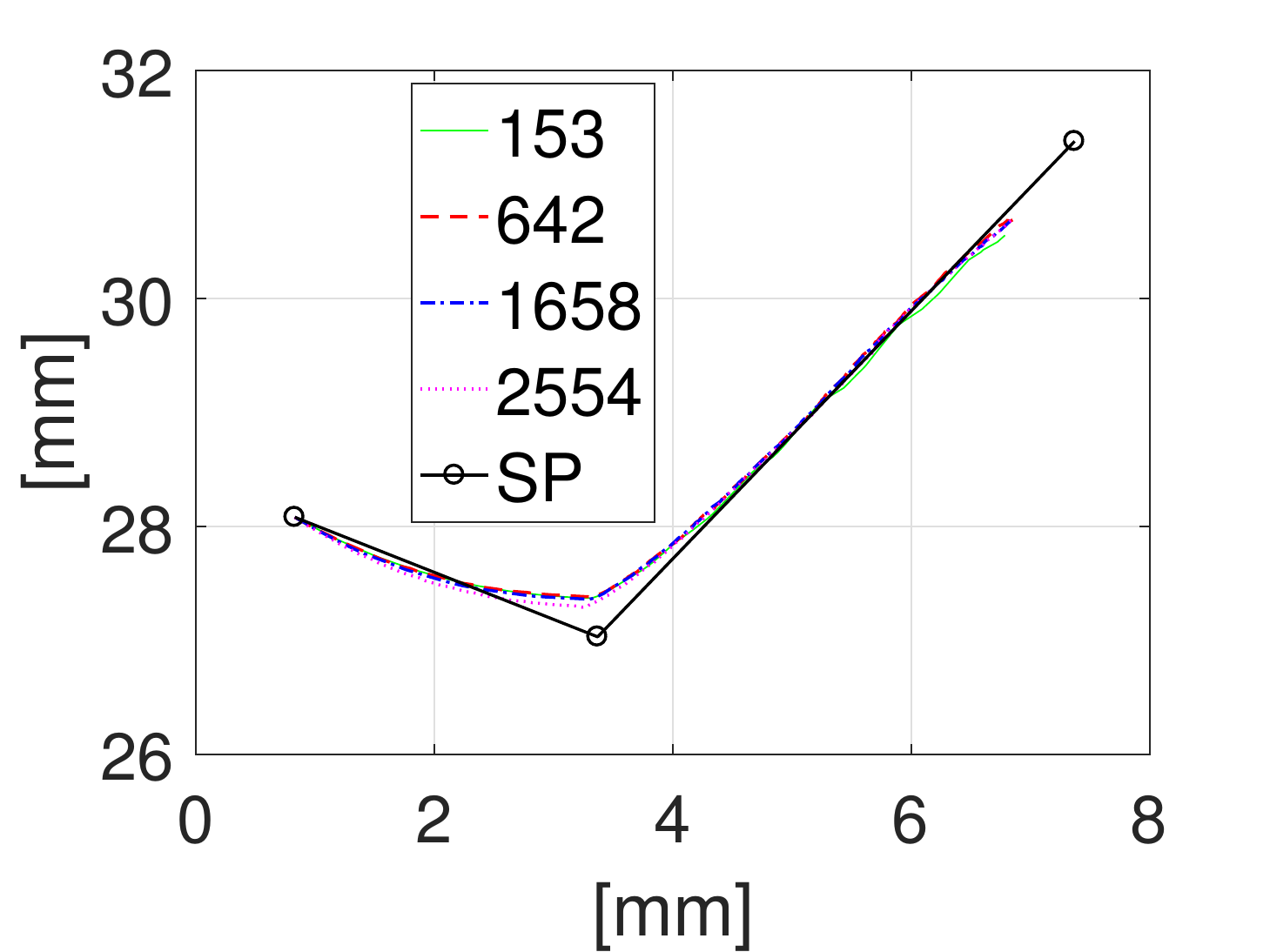}
		\caption{CCM I}
		\label{fig:SM1_paths}			
	\end{subfigure}  
	\begin{subfigure}{0.45\textwidth}
		\centering
		\includegraphics[scale = 0.45]{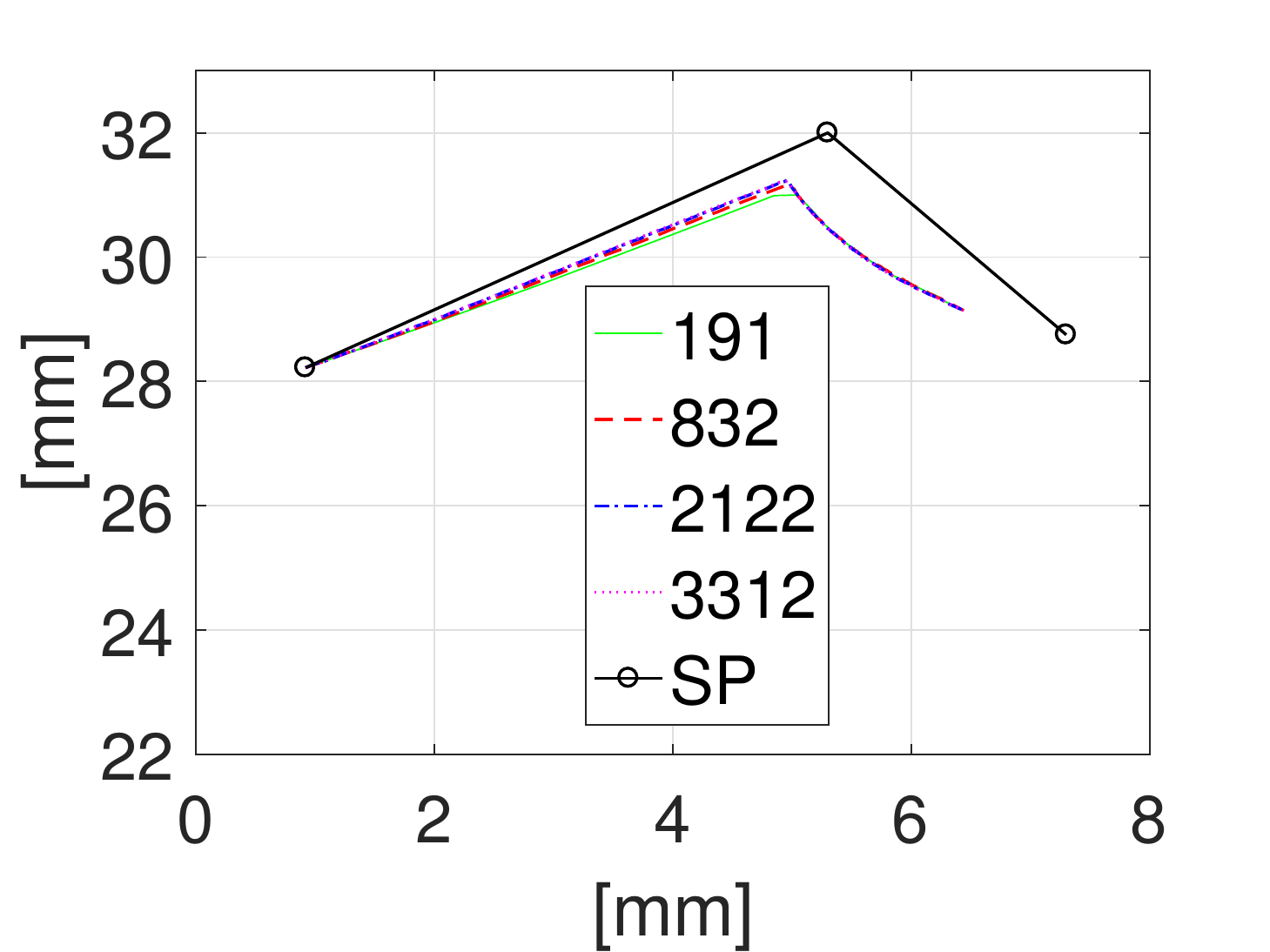}
		\caption{CCM II} 
		\label{fig:SM2_paths}			
	\end{subfigure}
	\begin{subfigure}{0.45\textwidth}
		\centering
		\includegraphics[scale = 0.5]{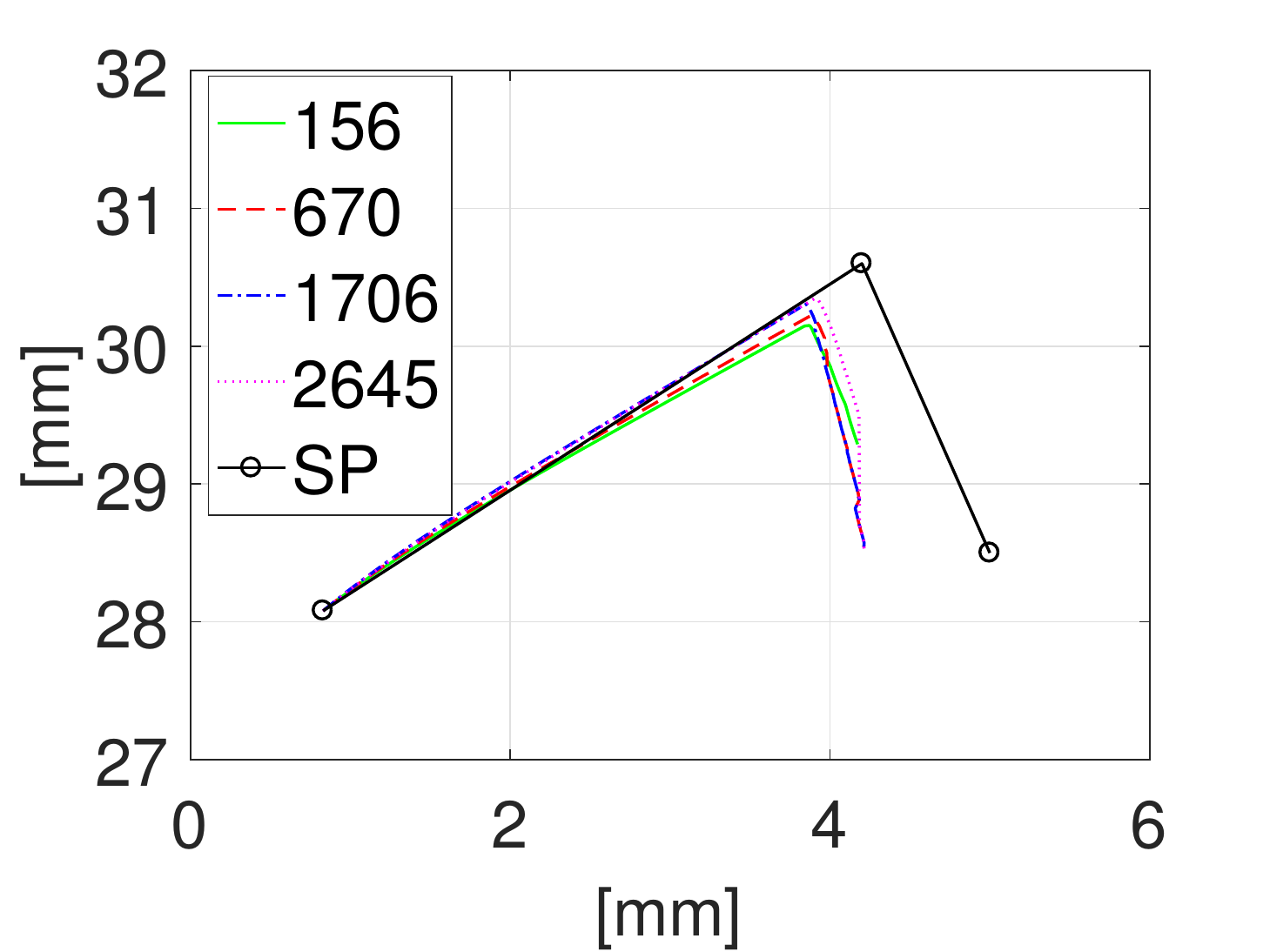}
		\caption{CCM III} 
		\label{fig:S_paths}			
	\end{subfigure}
~\qquad\quad
	\begin{subfigure}{0.45\textwidth}
		\centering
		\includegraphics[scale = 0.5]{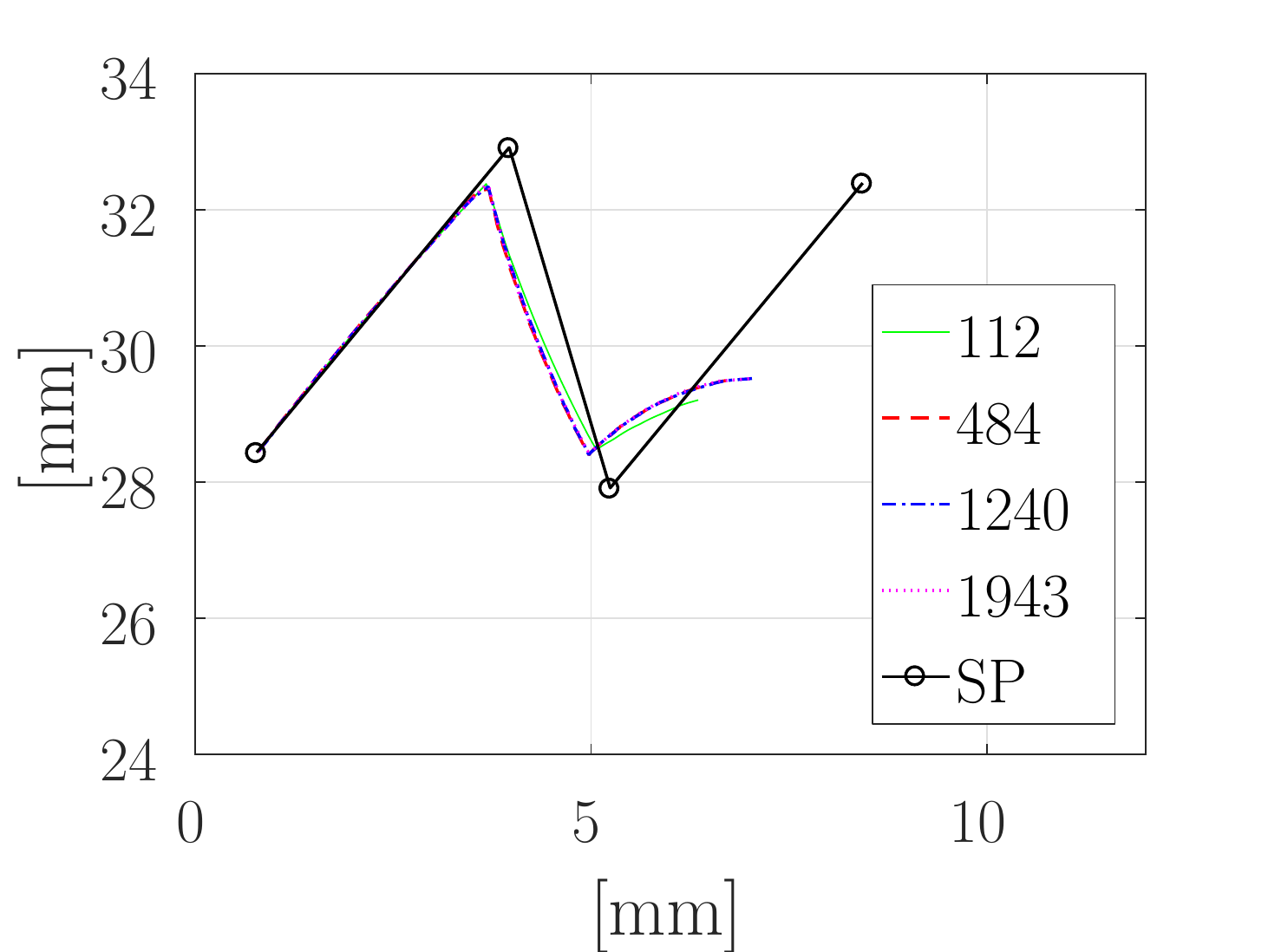}
		\caption{CCM IV} 
		\label{fig:Z_paths}			
	\end{subfigure}
	\caption{Paths traced by CCMs I-IV meshed with different number of hexagonal cells (inset). In fig. (\subref{fig:Z_paths}), desired path is rotated for better comparison. $SP$ represents the desired path.}	
	\label{fig:10} 
\end{figure}

Two different discretization types exist in the problem --- that of the design domain for the analysis using $n_{el}$ hexagonal finite elements, and the other of the design domain for the optimization using $n_m$ circular masks. Here, we examine the convergence behavior for refinement of the analysis discretization, keeping the number of masks fixed. The primary characteristics (overall shape and desired kinks) of the desired output displacements are captured well (Table \ref{T3}). Fig. \ref{fig:10} shows paths traced by the CCMs with different mesh resolutions with hexagonal cells. The green paths depict those obtained with the original resolution used in the synthesis. Errors in shapes ($A_\mathrm{err}$, $B_\mathrm{err}$) and sizes of the respective paths (Fig. \ref{fig:7}) are depicted in Table \ref{T6}. With mesh refinement, $\zeta_l$ exhibits a decreasing trend for all CCMs. This is expected since with increased number of elements, relatively more flexible CCM is simulated resulting in path lengths being closer to those desired. $A_\mathrm{err}$, $B_\mathrm{err}$ decrease in case of CCMs I and II. For CCM III, $A_\mathrm{err}$ increases marginally though there is a sharp initial drop in $B_\mathrm{err}$ and $\zeta_l$. Shape errors do not change much for CCM IV. With mesh refinement, paths predicted by the resulting CCMs are marginally more proximal, primarily in size, to the respective desired paths than those obtained via CCMs with original resolution. No significant change in shape is observed. Except in case of CCM III, not much improvement is observed with mesh refinement suggesting that coarse meshes are quite adequate for use in synthesizing CCMs. Use of refined meshes in synthesis is preferrable, albeit, with increase in computational cost. With the possibility of modeling a candidate CCM continua with refined hexagonal meshes, availability of better and efficient analyses methods, mesh dependence and imposition of minimum length scales \cite{guest2004achieving} with hexagonal cells can be explored.

\begin{table}
	\centering
	\caption{$A_\mathrm{err}$, $B_\mathrm{err}$, $\zeta_l$ and $\theta_\mathrm{diff}$ measures for the paths traced via finer meshes with respect to respective specified paths}
	\label{T6} 
	\resizebox{\columnwidth}{!}{%
		\begin{tabular}{ |c|c|c|c|c|c|}
			\hline
			\textbf{Mechanisms} & \textbf{CCMs (\# cells)}&  $A_\mathrm{err}$ & $B_\mathrm{err}$&\textbf{$\zeta_l$} (\%) &$\theta_\mathrm{diff}$(degree)  \\ \hline
			\multirow{4}{*}{CCM I } & (153)& 0.0613 & 0.0750 &13.5282 & 6.2848 \\
			& (642)&0.0374&	0.0474	&12.6154&	7.5042\\
			& (1658)&0.0336&	0.0494&	12.7506&	7.8421\\
			& (2554)&0.0262&	0.0287&	11.8527&	8.5042\\ \hline
			\multirow{3}{*}{CCM II } & (191)& 0.1879 & 0.1559 &19.6902& 7.2912 \\
			& (832)&0.1534&	0.1299&	19.1125	&7.1046 \\
			& (2122)&0.1423	&0.1198&	19.0071&	6.9247\\
			& (3312)&0.1418	&0.1202	&18.7344&	5.2570 \\ \hline
			\multirow{3}{*}{CCM III } & (156)& 0.0332 & 0.5312& 23.3727 & 4.9178\\
			& (670)&0.0820&	0.1197&	15.1757&	5.4855 \\
			& (1658)&0.0836	&0.1303&	13.5780&	5.9388  \\
			& (2554)&0.1007&	0.1587&	11.2807&	5.9825 \\ \hline
			\multirow{3}{*}{CCM IV } & (112)& 0.5486 & 0.5581& 30.1120 & 56.1630 \\
			& (484)& 0.5845&	0.5988&	27.8730&	56.9761 \\
			& (1240)&0.6000	&0.6215&	25.9711&	57.0080 \\
			& (1943)&0.5995&	0.6199&	25.1355&	57.5288\\ \hline
		\end{tabular}
	}
\end{table}

\subsection{Zero order search}
When synthesizing CCMs, to perform the contact analysis for each candidate continuum, non-existing cells/elements from the parent finite element mesh are removed (temporarily) and nodal displacements of the remnant continuum are computed. A candidate continuum changes in each iteration  and so does the set of removable elements. As displacement sensitivities cannot be computed at all nodes in the parent mesh, 
implementation of gradient based search becomes difficult. Many other factors, such as, non-convergence of large displacement analysis (which can stall gradient search) \cite{saxena2013combined}, material model being perfectly binary with one of the five mask (design) variables being discrete\footnote{center coordinates and radius of a mask are continuous design variables; the fourth variable that helps decide whether a contact surface within a mask exists, is discrete, and the fifth, that determines the radius of the rigid contact surface, is continuous.}, additionally leads us to employ a zero-order search. An evident drawback is the significantly large number of iterations required in many of which, CCM candidates get penalized for the absence of input forces, output port, fixed dofs and well-connectness, and thus do not get evaluated via the finite element contact analysis. Scalability and efficiency, important features of the search, are not within the scope of the manuscript as they require a separate and detailed study, intended in future. 

\subsection{Mask Addition and Deletion}
\label{MSKADDDEL}
Mask addition/deletion has been explored in previous efforts (e.g., \cite{saxena2011adaptive}, \cite{saxena2013combined}). Mask addition was performed by stochastically and precisely superposing some extra masks over some already existing masks so that the previous topology remained unperturbed. Parameters of the newly added masks were mutated in the subsequent step to generate a new topology which was evaluated. Mask deletion was performed by identifying those that were redundant, that is, they were either enclosed within other masks or, laid over the other previously processed masks. Such masks do not directly contribute in determination of the topology. However, in the context of CCM synthesis herein, they may still enclose active contact surfaces, e.g.,  CCMs I, II and IV (Figs. \ref{fig:7}a,\,c,\,g). As in \cite{saxena2011adaptive} and \cite{saxena2013combined}, mask addition and deletion can be performed stochastically (i.e., with some probability, just as the design variables are mutated) with the proposed approach as well, making it more generic so that topology optimization of CCMs with MMOS does not depend on the prescribed number of masks. In essence, the number of masks may be evolved simultaneously with the continuum topology. When performing mask deletion, those generating active contact surfaces may not be considered redundant and therefore may not be deleted.

\subsection{Boundary smoothing, Mean Value Coordinates and Computational Cost} \label{CMPCSTs}
Hexagonal cells, by virtue of their geometry, circumvent geometric singularities like the checkerboard patterns and point connections. However, they leave serrated boundaries \cite{saxena2011topology}. Boundary smoothing subdues these notches thereby making it easier for contact analysis to be accomplished. Boundaries still get approximated using piece-wise linear segments, however, alternating contact forces are avoided \cite{corbett2014nurbs}. For contact analysis to be more accurate, and especially for its implementation with friction, which will be explored in future work, the analysis will benefit from a discretization based on $C^1-$ continuous curves for which tangent and normal vectors at each point is well defined. 

\begin{figure}[h!]
	\centering
	\begin{subfigure}{0.45\columnwidth}
		\centering
		\includegraphics[scale=0.5]{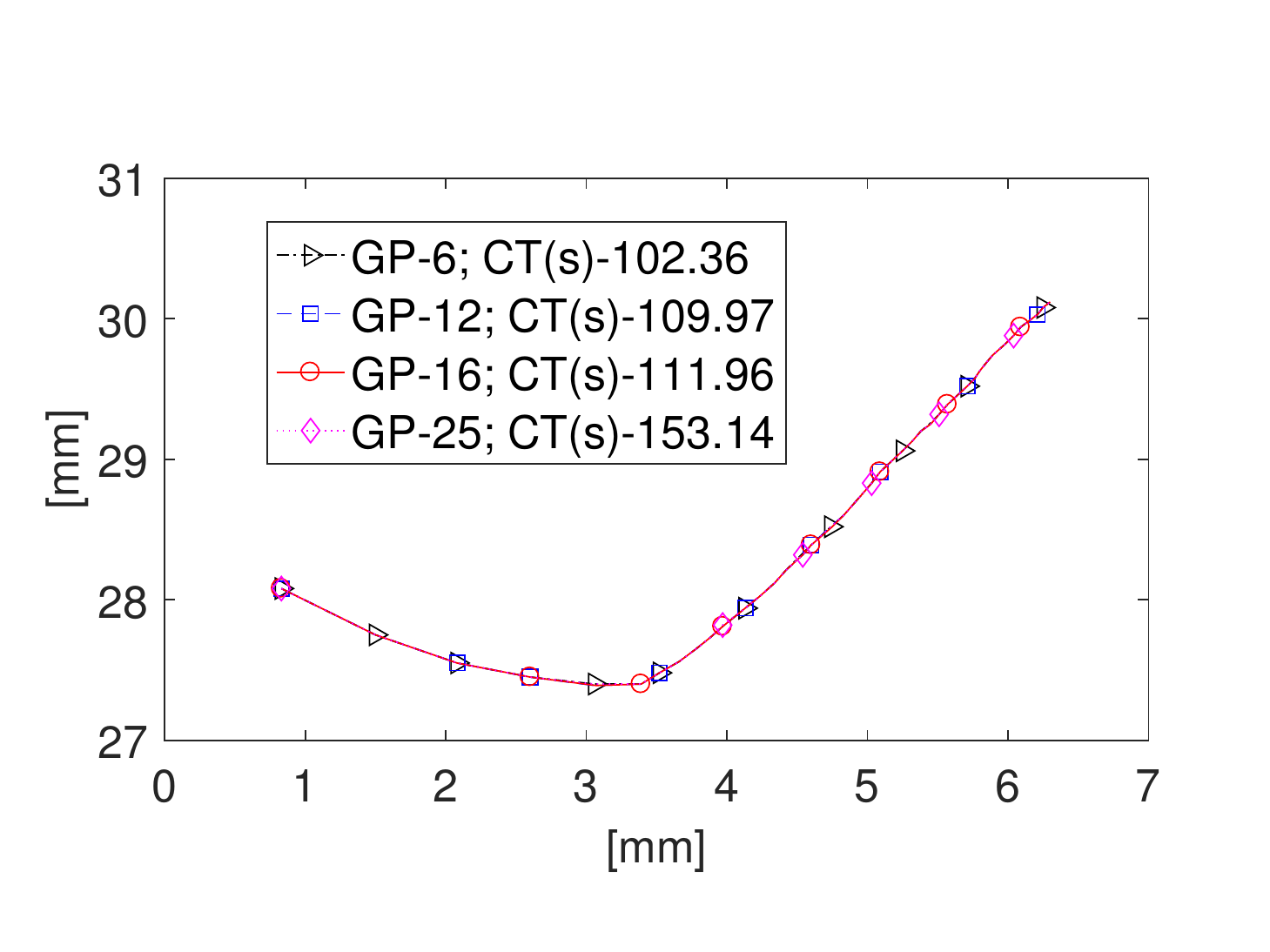}
		\caption{}
	\end{subfigure}
	\begin{subfigure}{0.45\columnwidth}
		\centering
		\includegraphics[scale=0.45]{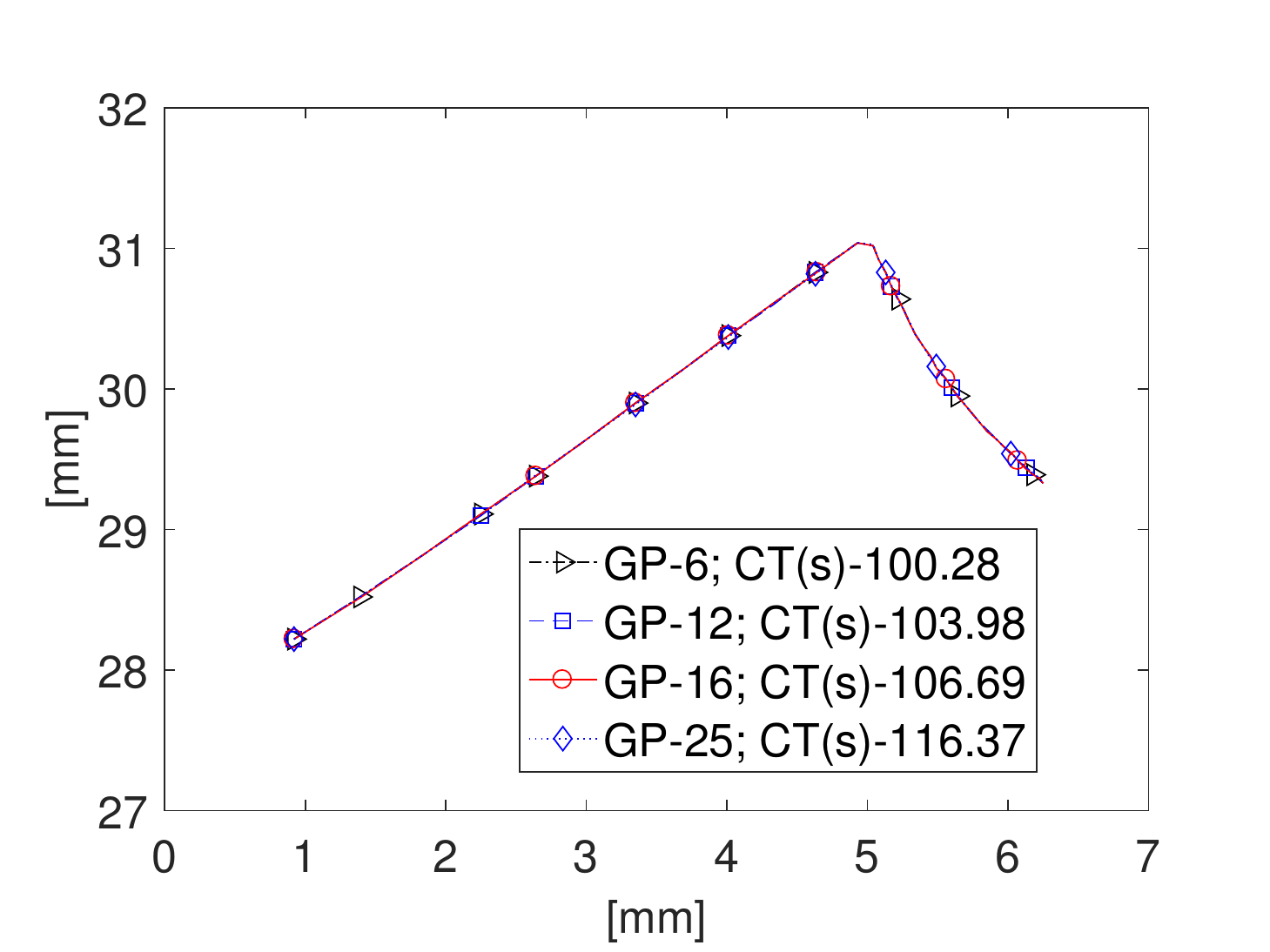}
		\caption{}
	\end{subfigure}
	\begin{subfigure}{0.45\columnwidth}
		\centering
		\includegraphics[scale=0.5]{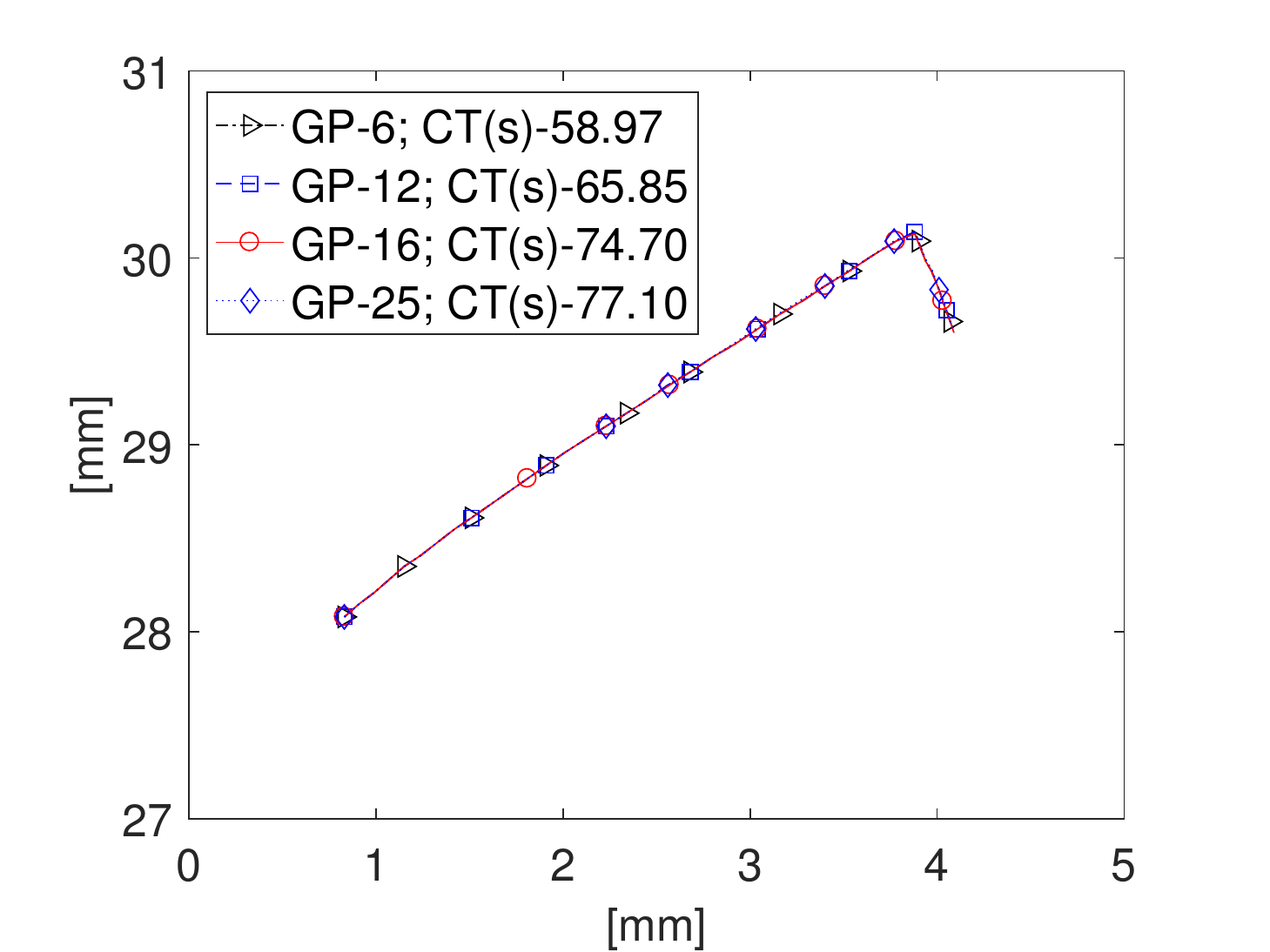}
		\caption{}
	\end{subfigure}
	\begin{subfigure}{0.45\columnwidth}
		\centering
		\includegraphics[scale=0.5]{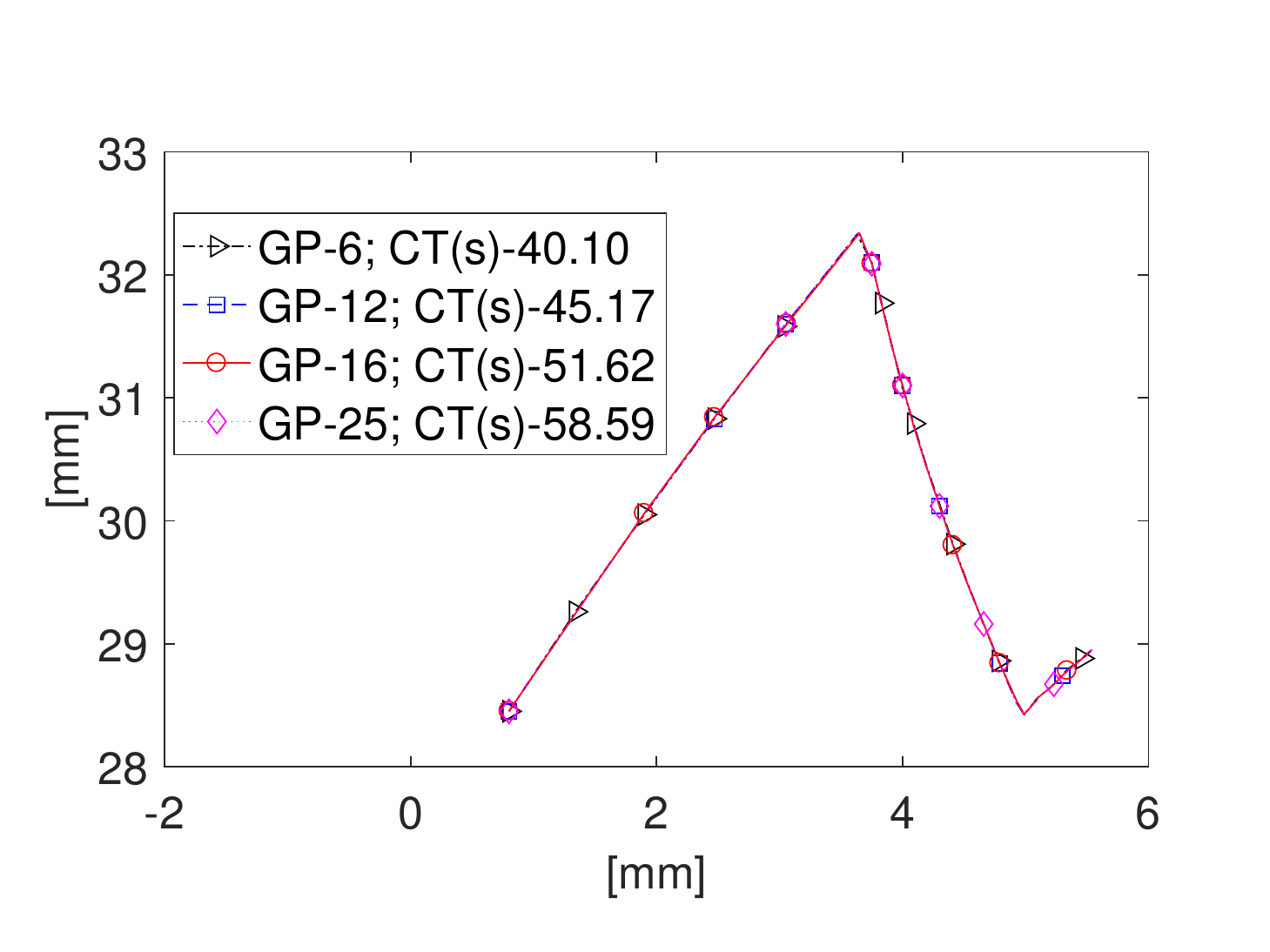}
		\caption{}
	\end{subfigure}
	\caption{Paths and CPU time with various Gauss points, \textbf{Key}:\, GP--Gauss Points, CT(s)--CPU time in seconds.}
	\label{fig:gaupt}
\end{figure}

Computationally, only few Gauss points are needed to solve the problem, but for accuracy and robustness of contact computations, especially in presence of concave cells resulting from boundary smoothing, 25 Gauss points per triangle of a hexagonal cell are employed in the finite elements analysis \cite{sukumar2004conforming}. We further analyze all CCMs, considering them as candidate compliant continua within the search process, with fewer Gauss points and provide an estimate of the relative CPU times. The output paths almost overlap (Fig. \ref{fig:gaupt}), with reduced CPU times suggesting that less number of Gauss points could be employed when performing integration. 

\begin{figure}[h!]
	\begin{subfigure}[t]{0.45\columnwidth}
		\centering
		\includegraphics[scale = 1.5]{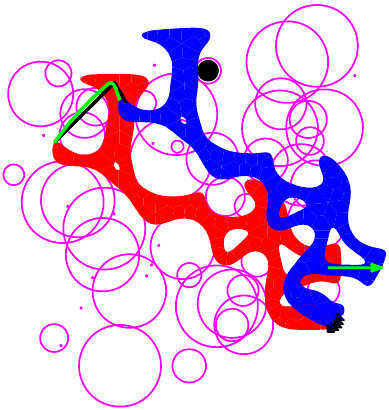}
		\caption{CCM after 6000 iterations} 
		\label{fig:fig_12a}			
	\end{subfigure}
	\begin{subfigure}[t]{0.45\columnwidth}
		\centering
		\includegraphics[scale = 0.4]{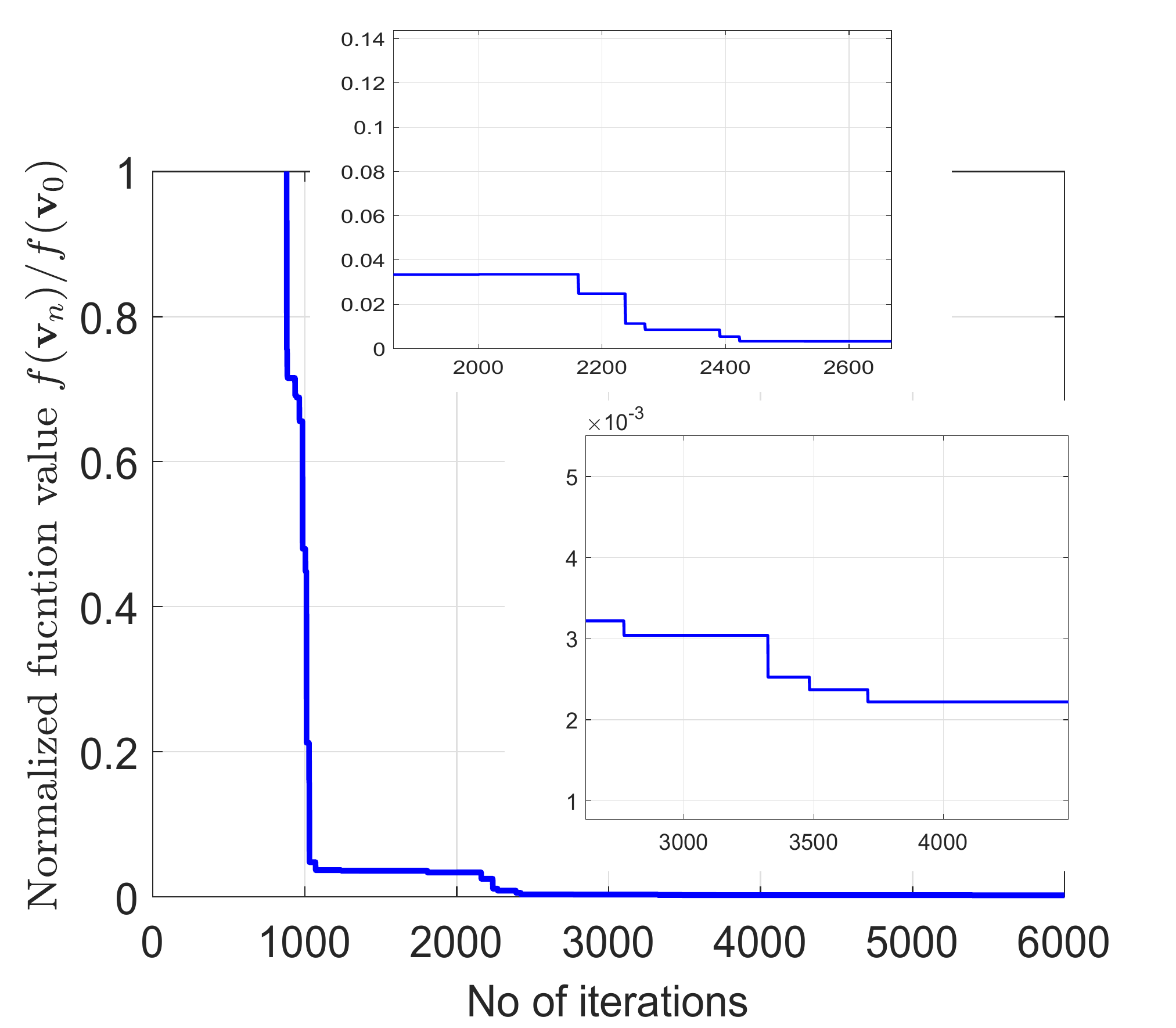}
		\caption{Convergence history} 
		\label{fig:fig_12b}			
	\end{subfigure}
	\caption{Computational cost and convergence history with INTEL CORE(TM) i5-6000 CPU @ 2.70 GHz. Solution (a) is obtained after $6000$ function evaluations. A feasible candidate continuum, that having the required input/output ports, some fixed boundaries, and that is well connected, is not available for the first 880 iterations. The overall synthesis takes about 12 hours.}	
	\label{fig:fig_12} 
\end{figure}

We report an estimate of the computational cost with the CCM presented in Fig.  \ref{fig:fig_12a} by synthesizing it on INTEL CORE(TM) i5-6000 CPU @ 2.70 GHz machine. The associated converge history is depicted in Fig. \ref{fig:fig_12b}. Each optimization iteration can be categorized into two main steps. In the first step, (i) one performs systematic mutation of the design variables using the hill-climber approach according to section \ref{HC}, (ii)  examines for the presence of input node, output node, some fixed boundary and removes dangling elements in the candidate solution, and (iii) implements boundary smoothing. The first step is repeated until one gets a potential continuum for the second stage (see flow chart in Fig. \ref{fig:5}). In the second stage, the N-R solution procedure is performed and thereafter, the FSD objective is evaluated. In each N-R iteration, (i) detection of contact pair, (ii) calculation of contact forces and matrices, and (iii) execution of the finite element analysis, are performed. For a potential continuum, the second step uses substantially more computational time. The first step may require many iterations to yield a feasible CCM with zero order search leading to significant contribution to the computational cost (Fig. \ref{fig:fig_12b}), which is a shortcoming of the synthesis method. It takes close to $880$ iterations (Fig. \ref{fig:fig_12b}) for the first feasible CCM to appear. For the example presented in Fig. \ref{fig:fig_12a}, the overall search culminates approximately within 12 hours. Note that the solution is different from that in Fig. \ref{fig:7}c though both are synthesized using identical design specifications. Mankame and Ananthasuresh \cite{mankame2007synthesis}, p. 2594, report the required computational time close to 72 hours (266 MHz, single processor, Sun Ultra Sparc 5 workstation) with the gradient based search, beams representing the design space and contact locations prespecified. 

\subsection{Force transfer and Failure-free CCMs} \label{CCMFFT}
Contact-aided compliant mechanisms are synthesized herein to primarily demonstrate that the proposed approach is capable of achieving kinematic intricacies (single or multiple kinks) in path generation. While force transfer and/or attaining compliant continua that does not fail is not the main goal, force transfer could be achieved by incorporating a spring of suitable stiffness, constant or non-constant, at the output. The spring should cater to cases where there are abrupt changes in the path, and/or if the continuum indulges in motion transfer prior to force transfer. One also expects a monolithic continuum, contact-aided or otherwise, to not fail when performing a mechanical transfer task for a single or many cycles. Failure theories have been incorporated previously (e.g., \cite{canfield2007multi}) in topology design of compliant mechanisms. A failure-free objective can either be addressed along with the kinematic requirement in the multi-objective setting \cite{canfield2007multi}, or separately and once the kinematic objective is achieved. In the former case, one could choose the best solution from a Pareto-front, and in the latter, noting that the design space for CCMs can be non-convex, one could further evolve multiple solutions (e.g., Figs. \ref{fig:8b}, \ref{fig:8c} and \ref{fig:fig_12a}), all satisfying the kinematic requirement, to ensure they sustain a prescribed number of cycles. 

\subsection{Presence of numerous rigid contact surfaces}
In case of  CCMs I, II, and IV,  numerous rigid contact surfaces are generated by the synthesis approach though only a few interact actively (Fig. \ref{fig:7}), not necessarily at the same time. Presence of inactive contact surfaces contributes to the computational cost of detecting contact pairs. However, their presence may help when CCMs are sought for more complex tasks, e.g., tracing output paths with more than two kinks, or attaining intricate shape profiles in case of motion generation applications. In some cases, suitable sizes, shapes and positioning of these \textquoteleft external' contact regions may cause some members of the CCM to buckle, a phenomenon that may be utilized for applications involving, say, static balancing. Design of CCMs for many such applications will be explored in future.

\section{Closure} \label{closure}
A continuum synthesis approach for large deformation, path generating Contact-aided Compliant Mechanisms is presented. Honeycomb tessellation is employed to represent the design region and negative circular masks for material assignment, and also, to suspend rigid contact surfaces. The novelty of the proposed method is that it captures both, \textit{self} and \textit{mutual} contact modes. The synthesis process is exemplified with four path-generating CCMs, each tracing a large output path with at least one non-differentiable point. A single-piece CCM tracing a \textquoteleft Z\textquoteright\, path, that has two kinks, exemplifies the capability of the proposed approach to capture multiple self/mutual contact modes to yield intricate deformation profiles.  In all four cases, path shapes characterized by $A_\mathrm{err}\,\text{and}\, B_\mathrm{err}$ are captured well. Discrepancies in the path lengths and orientations exist,  as expected, due to lower weights employed in the objective and due to a finite number of masks. Coarse meshes and use of fewer Gauss points in the analysis seem adequate in capturing the desired kinematic characteristics by and large, however with fine meshes, accuracy does improve. As a stochastic search is employed, computational costs are high. Future endeavors will be directed towards making analysis and search procedures more efficient by considering the possibility of using first and second order searches, implementing friction in synthesis, and exploring CCM design for applications like static balancing.

\bibliography{myreference}

\begin{thebibliography}{10}

\bibitem{mankame2004topology}
N.~D. Mankame and G.~Ananthasuresh, ``Topology optimization for synthesis of
  contact-aided compliant mechanisms using regularized contact modeling,'' {\em
  Computers \& structures}, vol.~82, no.~15, pp.~1267--1290, 2004.

\bibitem{kumar2016synthesis}
P.~Kumar, R.~A. Sauer, and A.~Saxena, ``Synthesis of $c^0$ path-generating
  contact-aided compliant mechanisms using the material mask overlay method,''
  {\em Journal of Mechanical Design}, vol.~138, no.~6, p.~062301, 2016.

\bibitem{wriggers2006computational}
P.~Wriggers, {\em Computational contact mechanics}, vol.~30167.
\newblock Springer, Berlin Heidelberg, 2006.

\bibitem{howell2001compliant}
L.~L. Howell, {\em Compliant mechanisms}.
\newblock John Wiley \& Sons, New York, 2001.

\bibitem{ananthasuresh1994methodical}
G.~Ananthasuresh, S.~Kota, and Y.~Gianchandani, ``A methodical approach to the
  design of compliant micromechanisms,'' in {\em Solid-state sensor and
  actuator workshop}, vol.~1994, pp.~189--192, SC: IEEE, 1994.

\bibitem{nishiwaki1998topology}
S.~Nishiwaki, M.~I. Frecker, S.~Min, and N.~Kikuchi, ``Topology optimization of
  compliant mechanisms using the homogenization method,'' {\em International
  Journal for numerical methods in engineering}, vol.~42, no.~3, pp.~535--559,
  1998.

\bibitem{frecker1997topological}
M.~Frecker, G.~Ananthasuresh, S.~Nishiwaki, N.~Kikuchi, and S.~Kota,
  ``Topological synthesis of compliant mechanisms using multi-criteria
  optimization,'' {\em Journal of Mechanical design}, vol.~119, no.~2,
  pp.~238--245, 1997.

\bibitem{saxena2000optimal}
A.~Saxena and G.~Ananthasuresh, ``On an optimal property of compliant
  topologies,'' {\em Structural and multidisciplinary optimization}, vol.~19,
  no.~1, pp.~36--49, 2000.

\bibitem{sigmund1997design}
O.~Sigmund, ``On the design of compliant mechanisms using topology
  optimization,'' {\em Journal of Structural Mechanics}, vol.~25, no.~4,
  pp.~493--524, 1997.

\bibitem{saxena2001topology}
A.~Saxena and G.~Ananthasuresh, ``Topology synthesis of compliant mechanisms
  for nonlinear force-deflection and curved path specifications,'' {\em Journal
  of Mechanical Design}, vol.~123, no.~1, pp.~33--42, 2001.

\bibitem{pedersen2001topology}
C.~B. Pedersen, T.~Buhl, and O.~Sigmund, ``Topology synthesis of
  large-displacement compliant mechanisms,'' {\em International Journal for
  numerical methods in engineering}, vol.~50, no.~12, pp.~2683--2705, 2001.

\bibitem{saxena2005synthesis}
A.~Saxena, ``Synthesis of compliant mechanisms for path generation using
  genetic algorithm,'' {\em Journal of Mechanical Design}, vol.~127, no.~4,
  pp.~745--752, 2005.

\bibitem{swan2004design}
C.~C. Swan and S.~F. Rahmatalla, ``Design and control of path-following
  compliant mechanisms,'' in {\em ASME 2004 International Design Engineering
  Technical Conferences and Computers and Information in Engineering
  Conference}, pp.~1173--1181, American Society of Mechanical Engineers, 2004.

\bibitem{ullah1997optimal}
I.~Ullah and S.~Kota, ``Optimal synthesis of mechanisms for path generation
  using fourier descriptors and global search methods,'' {\em Journal of
  Mechanical Design}, vol.~119, no.~4, pp.~504--510, 1997.

\bibitem{zahn1972fourier}
C.~T. Zahn and R.~Z. Roskies, ``Fourier descriptors for plane closed curves,''
  {\em Computers, IEEE Transactions on}, vol.~100, no.~3, pp.~269--281, 1972.

\bibitem{rai2007synthesis}
A.~K. Rai, A.~Saxena, and N.~D. Mankame, ``Synthesis of path generating
  compliant mechanisms using initially curved frame elements,'' {\em Journal of
  Mechanical Design}, vol.~129, no.~10, pp.~1056--1063, 2007.

\bibitem{rai2010unified}
A.~K. Rai, A.~Saxena, and N.~D. Mankame, ``Unified synthesis of compact planar
  path-generating linkages with rigid and deformable members,'' {\em Structural
  and Multidisciplinary Optimization}, vol.~41, no.~6, pp.~863--879, 2010.

\bibitem{saxena2008material}
A.~Saxena, ``A material-mask overlay strategy for continuum topology
  optimization of compliant mechanisms using honeycomb discretization,'' {\em
  Journal of Mechanical Design}, vol.~130, p.~082304, 2008.

\bibitem{saxena2011adaptive}
A.~Saxena, ``An adaptive material mask overlay method: modifications and
  investigations on binary, well connected robust compliant continua,'' {\em
  Journal of Mechanical Design}, vol.~133, p.~041004, 2011.

\bibitem{saxena2011topology}
A.~Saxena, ``Topology design with negative masks using gradient search,'' {\em
  Structural and Multidisciplinary Optimization}, vol.~44, no.~5, pp.~629--649,
  2011.

\bibitem{saxena2013combined}
A.~Saxena and R.~A. Sauer, ``Combined gradient-stochastic optimization with
  negative circular masks for large deformation topologies,'' {\em
  International Journal for Numerical Methods in Engineering}, vol.~93, no.~6,
  pp.~635--663, 2013.

\bibitem{mankame2002contact}
N.~D. Mankame and G.~Ananthasuresh, ``Contact aided compliant mechanisms:
  concept and preliminaries,'' in {\em ASME 2002 International Design
  Engineering Technical Conferences and Computers and Information in
  Engineering Conference}, pp.~109--121, American Society of Mechanical
  Engineers, 2002.

\bibitem{mankame2007synthesis}
N.~Mankame and G.~Ananthasuresh, ``Synthesis of contact-aided compliant
  mechanisms for non-smooth path generation,'' {\em International Journal for
  Numerical Methods in Engineering}, vol.~69, no.~12, pp.~2564--2605, 2007.

\bibitem{reddy2012systematic}
B.~V. S.~N. Reddy, S.~V. Naik, and A.~Saxena, ``Systematic synthesis of large
  displacement contact-aided monolithic compliant mechanisms,'' {\em Journal of
  Mechanical Design}, vol.~134, no.~1, p.~011007, 2012.

\bibitem{tummala2014design}
Y.~Tummala, A.~Wissa, M.~Frecker, and J.~E. Hubbard, ``Design and optimization
  of a contact-aided compliant mechanism for passive bending,'' {\em Journal of
  Mechanisms and Robotics}, vol.~6, no.~3, p.~031013, 2014.

\bibitem{kumar2017implementation}
P.~Kumar, A.~Saxena, and R.~A. Sauer, ``Implementation of self contact in path
  generating compliant mechanisms,'' in {\em Microactuators and
  Micromechanisms}, pp.~251--261, Springer, Cham, 2017.

\bibitem{cannon2005compliant}
J.~R. Cannon and L.~L. Howell, ``A compliant contact-aided revolute joint,''
  {\em Mechanism and Machine Theory}, vol.~40, no.~11, pp.~1273--1293, 2005.

\bibitem{moon2007bio}
Y.-M. Moon, ``Bio-mimetic design of finger mechanism with contact aided
  compliant mechanism,'' {\em Mechanism and Machine Theory}, vol.~42, no.~5,
  pp.~600--611, 2007.

\bibitem{aguirre2008fabrication}
M.~Aguirre, G.~Hayes, M.~Frecker, J.~Adair, and N.~Antolino, ``Fabrication and
  design of a nanoparticulate enabled micro forceps,'' in {\em ASME 2008
  International Design Engineering Technical Conferences and Computers and
  Information in Engineering Conference}, pp.~361--370, American Society of
  Mechanical Engineers, 2008.

\bibitem{mehta2009stress}
V.~Mehta, M.~Frecker, and G.~A. Lesieutre, ``Stress relief in contact-aided
  compliant cellular mechanisms,'' {\em Journal of Mechanical Design},
  vol.~131, no.~9, p.~091009, 2009.

\bibitem{saxena2013contact}
A.~Saxena, ``A contact-aided compliant displacement-delimited gripper
  manipulator,'' {\em Journal of Mechanisms and Robotics}, vol.~5, no.~4,
  p.~041005, 2013.

\bibitem{calogero2016dynamic}
J.~Calogero, M.~Frecker, Z.~Hasnain, and J.~E. Hubbard~Jr, ``A dynamic spar
  numerical model for passive shape change,'' {\em Smart Materials and
  Structures}, vol.~25, no.~10, p.~104006, 2016.

\bibitem{sauer2015unbiased}
R.~A. Sauer and L.~De~Lorenzis, ``An unbiased computational contact formulation
  for 3d friction,'' {\em International Journal for Numerical Methods in
  Engineering}, vol.~101, no.~4, pp.~251--280, 2015.

\bibitem{saxena2003honeycomb}
R.~Saxena and A.~Saxena, ``On honeycomb parameterization for topology
  optimization of compliant mechanisms,'' in {\em ASME 2003 International
  Design Engineering Technical Conferences and Computers and Information in
  Engineering Conference}, pp.~975--985, American Society of Mechanical
  Engineers, 2003.

\bibitem{langelaar2007use}
M.~Langelaar, ``The use of convex uniform honeycomb tessellations in structural
  topology optimization,'' in {\em 7th world congress on structural and
  multidisciplinary optimization, Seoul, South Korea, May}, pp.~21--25, 2007.

\bibitem{saxena2007honeycomb}
R.~Saxena and A.~Saxena, ``On honeycomb representation and sigmoid material
  assignment in optimal topology synthesis of compliant mechanisms,'' {\em
  Finite Elements in Analysis and Design}, vol.~43, no.~14, pp.~1082--1098,
  2007.

\bibitem{talischi2009honeycomb}
C.~Talischi, G.~H. Paulino, and C.~H. Le, ``Honeycomb wachspress finite
  elements for structural topology optimization,'' {\em Structural and
  Multidisciplinary Optimization}, vol.~37, no.~6, pp.~569--583, 2009.

\bibitem{talischi2012polytop}
C.~Talischi, G.~H. Paulino, A.~Pereira, and I.~F. Menezes, ``Polytop: a matlab
  implementation of a general topology optimization framework using
  unstructured polygonal finite element meshes,'' {\em Structural and
  Multidisciplinary Optimization}, vol.~45, no.~3, pp.~329--357, 2012.

\bibitem{saxena2010adaptive}
A.~Saxena, ``On an adaptive mask overlay topology synthesis method,'' in {\em
  ASME 2010 International Design Engineering Technical Conferences and
  Computers and Information in Engineering Conference}, pp.~675--684, American
  Society of Mechanical Engineers, 2010.

\bibitem{talischi2012polymesher}
C.~Talischi, G.~H. Paulino, A.~Pereira, and I.~F. Menezes, ``Polymesher: a
  general-purpose mesh generator for polygonal elements written in matlab,''
  {\em Structural and Multidisciplinary Optimization}, vol.~45, no.~3,
  pp.~309--328, 2012.

\bibitem{talischi2010polygonal}
C.~Talischi, G.~H. Paulino, A.~Pereira, and I.~F. Menezes, ``Polygonal finite
  elements for topology optimization: a unifying paradigm,'' {\em International
  Journal for Numerical Methods in Engineering}, vol.~82, no.~6, pp.~671--698,
  2010.

\bibitem{kumar2015topology}
P.~Kumar and A.~Saxena, ``On topology optimization with embedded boundary
  resolution and smoothing,'' {\em Structural and Multidisciplinary
  Optimization}, vol.~52, no.~6, pp.~1135--1159, 2015.

\bibitem{corbett2014nurbs}
C.~J. Corbett and R.~A. Sauer, ``Nurbs-enriched contact finite elements,'' {\em
  Computer Methods in Applied Mechanics and Engineering}, vol.~275, pp.~55--75,
  2014.

\bibitem{kumarembedded}
P.~Kumar and A.~Saxena, ``On embedded recursive boundary smoothing in topology
  optimization with polygonal mesh and negative masks,'' {\em AMM india, IIT
  Roorkee}, pp.~568--575, 2013.

\bibitem{floater2003mean}
M.~S. Floater, ``Mean value coordinates,'' {\em Computer Aided Geometric
  Design}, vol.~20, no.~1, pp.~19--27, 2003.

\bibitem{hormann2006mean}
K.~Hormann and M.~S. Floater, ``Mean value coordinates for arbitrary planar
  polygons,'' {\em ACM Transactions on Graphics (TOG)}, vol.~25, no.~4,
  pp.~1424--1441, 2006.

\bibitem{sukumar2004conforming}
N.~Sukumar and A.~Tabarraei, ``Conforming polygonal finite elements,'' {\em
  International Journal for Numerical Methods in Engineering}, vol.~61, no.~12,
  pp.~2045--2066, 2004.

\bibitem{sukumar2006recent}
N.~Sukumar and E.~Malsch, ``Recent advances in the construction of polygonal
  finite element interpolants,'' {\em Archives of Computational Methods in
  Engineering}, vol.~13, no.~1, pp.~129--163, 2006.

\bibitem{zienkiewicz2005finite}
O.~C. Zienkiewicz and R.~L. Taylor, {\em The finite element method for solid
  and structural mechanics}.
\newblock Elsevier, Butterworth-Heinemann, 2005.

\bibitem{wriggers2008nonlinear}
P.~Wriggers, {\em Nonlinear finite element methods}.
\newblock Springer Science \& Business Media, Berlin Heidelberg, 2008.

\bibitem{kumar2017_diss}
{Kumar, Prabhat}, {\em {Synthesis of Large Deformable Contact-Aided Compliant
  Mechanisms using Hexagonal cells and Negative Circular Masks}}.
\newblock PhD thesis, {Indian Institute of Technology Kanpur}, {2017}.

\bibitem{knuth1998art}
D.~E. Knuth, {\em The art of computer programming: sorting and searching},
  vol.~3.
\newblock Pearson Education, Upper Saddle River, New Jersey, 1998.

\bibitem{Russell:2003:AIM:773294}
S.~J. Russell and P.~Norvig, {\em Artificial Intelligence: A Modern Approach}.
\newblock Pearson Education, Upper Saddle River, New Jersey, 2~ed., 2003.

\bibitem{tikhonov2013numerical}
A.~N. Tikhonov, A.~Goncharsky, V.~Stepanov, and A.~G. Yagola, {\em Numerical
  methods for the solution of ill-posed problems}, vol.~328.
\newblock Springer Science \& Business Media, Dordrecht, 2013.

\bibitem{kumar2015synthesis}
P.~Kumar, R.~A. Sauer, and A.~Saxena, ``On synthesis of contact aided compliant
  mechanisms using the material mask overlay method,'' in {\em ASME 2015
  International Design Engineering Technical Conferences and Computers and
  Information in Engineering Conference}, pp.~V05AT08A017--V05AT08A017,
  American Society of Mechanical Engineers, 2015.

\bibitem{guest2004achieving}
J.~K. Guest, J.~Pr{\'e}vost, and T.~Belytschko, ``Achieving minimum length
  scale in topology optimization using nodal design variables and projection
  functions,'' {\em International Journal for Numerical Methods in
  Engineering}, vol.~61, no.~2, pp.~238--254, 2004.

\bibitem{canfield2007multi}
S.~L. Canfield, D.~L. Chlarson, A.~Shibakov, J.~D. Richardson, and A.~Saxena,
  ``Multi-objective optimization of compliant mechanisms including failure
  theories,'' in {\em ASME 2007 International Design Engineering Technical
  Conferences and Computers and Information in Engineering Conference},
  pp.~179--190, American Society of Mechanical Engineers, 2007.

\end{thebibliography}
\bibliographystyle{hieeetr}
 \end{document}